\documentclass[]{jfm}
\usepackage{graphicx}
\usepackage{caption}
\usepackage{subcaption}
\usepackage{float}
\usepackage{newtxtext}
\usepackage{newtxmath}
\usepackage{natbib}
\usepackage{hyperref}
\usepackage[final]{changes}

\hypersetup{
    colorlinks = true,
    urlcolor   = blue,
    citecolor  = black,
}

\newcommand{\RomanNumeralCaps}[1]

\title{Computational Modeling and Analysis of the Coupled Aero-Structural Dynamics in Bat‑Inspired Wings}

\author{Sushrut Kumar\aff{1},
  Jung-Hee Seo\aff{1}
 \and Rajat Mittal\aff{1}\corresp{\email{mittal@jhu.edu}}}

\affiliation{\aff{1}Department of Mechanical Engineering, Johns Hopkins University, Baltimore, MD, USA}

\begin{document}
\maketitle

\begin{abstract}
We employ a novel computational modeling framework to perform high-fidelity direct numerical simulations of aero-structural interactions in bat-inspired membrane wings. The wing of a bat consists of an elastic membrane supported by a highly articulated skeleton, enabling localized control over wing movement and deformation during flight. By modeling these complex deformations, along with realistic wing movements and interactions with the surrounding airflow, we expect to gain new insights into the performance of these unique wings. Our model achieves a high degree of realism by incorporating experimental measurements of the skeleton's joint movements to guide the fluid-structure interaction simulations. The simulations reveal that different segments of the wing undergo distinct aeroelastic deformations, impacting flow dynamics and aerodynamic loads. Specifically, the simulations show significant variations in the effectiveness of the wing in generating lift, drag, and thrust forces across different segments and regions of the wing. We employ a force partitioning method to analyze the causality of pressure loads over the wing, demonstrating that vortex-induced pressure forces are dominant while added mass contributions to aerodynamic loads are minimal. This approach also elucidates the role of various flow structures in shaping pressure distributions. Finally, we compare the fully articulated, flexible bat wing to equivalent stiff wings derived from the same kinematics, demonstrating the critical impact of wing articulation and deformation on aerodynamic efficiency.
\end{abstract}

\begin{keywords}
Authors should not enter keywords on the manuscript, as these must be chosen by the author during the online submission process and will then be added during the typesetting process (see \href{https://www.cambridge.org/core/journals/journal-of-fluid-mechanics/information/list-of-keywords}{Keyword PDF} for the full list). Other classifications will be added at the same time.
\end{keywords}

\section{Introduction}
\label{sec:intro} 
The aerodynamics of biological and bioinspired wings have been of great interest to the engineering community for many decades, but recent interest has been spurred by the development of micro-aerial vehicles (MAVs). Micro-aerial vehicles can be broadly classified into fixed-wing, rotating-wing, and flapping-wing MAVs. Flying animals such as insects, bats, and birds are being studied extensively to draw inspiration to design and develop bio-inspired MAVs \citep{alexander2002nature,azuma2012biokinetics,shyy2013introduction}. Natural fliers are capable of strokes with their wings using a complex blend of pitching, flapping, variations in stroke plane, wing area, camber, and sectional twist \citep{azuma2012biokinetics,shyy2013introduction}. Furthermore, unlike aircraft wings, the wings of natural fliers undergo significant deformations due to aerodynamic forces. Understanding this coupled fluid and structural dynamics involved in the locomotion of various natural fliers can provide useful design ideas to help build and efficiently operate these systems. Thus, the flapping flight of these flying animals presents a challenge and an opportunity for aerodynamic investigations. 

Bats are the only extant mammals that employ powered flight, and these animals have undergone millions of years of evolution to develop a flight apparatus that has allowed them to become agile fliers \citep{kunz2003bat}. The wing of a bat is fundamentally different from the wing of a bird or insect as it employs a highly deformable elastic membrane stretched upon the hand skeleton with elongated bones that undergoes a high degree of articulation and geometric deformation during flight \citep{norberg1987ecological,hedenstrom2015bat,swartz2015advances}. The skeletal muscles and bones impart flapping movement to the wing and also provide a highly localized control over wing pitch, twist and camber \citep{hedenstrom2009bird}.  The soft membrane wing can undergo aeroelastic phenomena such as area expansion and flutter when interacting with the complex flow field around it \citep{lauber2023rapid}. The interaction of airflow with the highly articulated and deformable membrane wing generates aerodynamic forces responsible for supporting the weight of the animal in flight and to enable complex flight maneuvers.  

The bat-hand wing in flapping flight offers an interesting and rich fluid-structure interaction problem. Although the biology of bats has been studied extensively, our knowledge about flight aerodynamics, specifically the coupled-aero-structural dynamics of the wing, is limited. The earliest studies on bat flight were mostly experimental and initially focused on the physiological and kinematics features \citep{hartman1963some,thomas1972physiology,norberg1987ecological,bullen2001bat,tian2006direct,riskin2008quantifying}. Later studies delved more into fluid dynamics using newer experimental techniques \citep{hedenstrom2007bat,muijres2008leading,hedenstrom2009bird,hubel2009time,hedenstrom2009high,johansson2010quantitative}. In these experimental studies, the lift and power were computed from the circulation and kinetic energy in the wake. The measurements from these experiments were limited to the wake due to the complexity involved in extracting flow data in the vicinity of the body and wing of the animal \citep{dabiri2005estimation,spedding2009piv,gutierrez2016lift}. 

Numerical simulations provide the opportunity to explore not just the fluid dynamics but the coupled physics of the fluid-structure interaction (FSI). Initial computational models of bat flight aerodynamics such as those by \cite{wang2015lift,wang2015numerical} along with Tafti and co-workers \citep{viswanath2014straight,sekhar2019canonical,windes2019determination,windes2020kinematic,rahman2022role} employed kinematics obtained from high-speed videogrammetry. These simulations provided interesting insights and data into the aerodynamics of bat wings but were one-way coupled and did not model the aero-structural dynamics. Jaiman and co-workers \citep{li2019novel,joshi2020full,joshi2020variational} were, to our knowledge, the first to employ an aeroelastic framework for simulating bat-inspired wings with bones, joints, and flexible membranes. Their simulations employed a comprehensive finite-element model for the structural dynamics but incorporated several simplifications into the wing kinematics such as prescribing the joint flapping motion using a sinusoidal rotation profile, and excluding the articulation of finger joints. The latter in particular, is a unique and important characteristic feature of bat wings.

Finally, the most recent work on bat wing aerodynamics is by \cite{lauber2023rapid} who focus on the modeling the coupled FSI of the outer section of the bat wing (i.e. the ``handwing''). Their model excludes the inner portion of the wing (i.e. the ``armwing'') which is comprised of the propatagium and the plagiopatagium. They employ a FSI model with a hyperelastic, anisotropic material model of the membrane, and match the movement of several points on the wing to the measurements of a bat in forward flight \citep{wolf2010kinematics}. However, similar to the previous work, they do not incorporate the articulation of the inner finger joints. They observed that isotropic membranes are most efficient before the onset of flutter and incorporating anisotropy into the material model delays flutter. However, as we will show, the armwing undergoes significant deformation during the flight and generates the vast majority of the aerodynamic load on the wing. Furthermore, the vortex structure from the handwing and the armwing interact with each other and therefore exclusion of the armwing has implications on the aerodynamics that are difficult to fully understand. 

We aim to perform FSI simulations of a bat wing in flapping flight with a high degree of realism with respect to the wing anatomy and kinematics, as well as the geometric and elastic deformation associated with active articulation and flow-induced deformation of the entire wing. These simulations are nominally based on experimental measurements of \cite{riskin2008quantifying}, and attempt to recapitulate the aero-structural dynamics of a bat in forward flight. Simpler models of the wings with reduced degrees-of-freedom are also simulated to extract the effect of wing deformability. The simulation results are subjected to a detailed analysis using the force partitioning method, which has been useful in providing insights into various vortex-dominated flows around wings like in \citep{zhang2015centripetal,menon2022contribution, zhu2023force}. These analyses provide insights into the mechanisms/features in the flow and the membrane dynamics that have dominant effects on the generation of lift, drag, and thrust, and the role that the unique properties of the bat membrane wing play in enabling flight.

\section{Computational Model}\label{sec:methods}
\subsection{Bat Wing Model}
\begin{figure}
    \centering
    \subfloat[ ]{\includegraphics[width=0.78\textwidth]{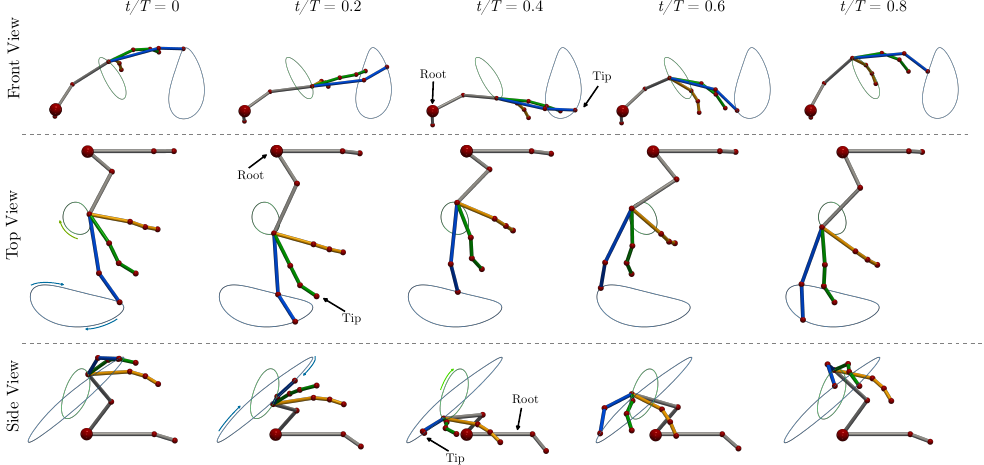}}
    \subfloat[ ]{\includegraphics[width=0.2\textwidth]{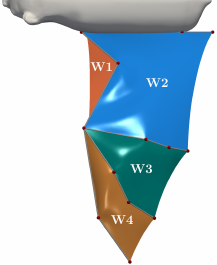}}
    \caption{Kinematics of the bat wing from \cite{riskin2008quantifying} that form the basis of the current model. (a) Three views of the wing skeleton during the flapping cycle including the trajectory of the wing-tip and the wrist joint. (b) Model of the wing planform adopted in the current study with the various segments identified as follows: W1:propatagium; W2:plagiopatagium; W3:dactylopatagium major; and W4:dactylopatagium medius }
    \label{fig:trajectory}
\end{figure}

A wing of a bat differs from the wings of other natural fliers such as insects and birds in significant ways. The wetted area of the wing is formed by the soft wing membrane, which is attached to the skeletal structure of the hand (the finger bones and joints), forming four major segments as shown in figure \ref{fig:trajectory}(b): the propatagium (indicated as ``W1''), the plagiopatagium (``W2''), the dactylopatagium major (``W3''), and the dactylopatagium medius (``W4''). There also exists the dactylopatagium minor, which is a fifth segment of the wing, but it has a very small surface area relative to the other segments and is therefore, neglected in the current study. 

The current model is based on the direct visualization and videogrammetry of a flying bat (\emph{Pteropus pumilus}) by \cite{riskin2008quantifying}. Figure \ref{fig:trajectory}(a) shows five snapshots of the skeletal structure of the bat and it can be seen that there is a high degree of active articulation during the flapping stroke as the digits move relative to the joints in order to change the shape of the wing. This results in very large geometric deformation in the wing during the flapping stroke. The hand wing of the bat may be considered to be divided into inelastic (the finger bones and joints that comprise the hand-wing skeleton) and elastic (membrane) components. The slender finger bones may undergo some bending deformation during flight, but the deformation is smaller compared to the elastic wing. We therefore treat the finger bones as inelastic in our model. In order to generate wing kinematics with a high degree of realism, we incorporate into the wing motion, the finger bone joint kinematics shown in figure \ref{fig:trajectory}(a). We have developed a pipeline that employs Fourier decomposition and cubic interpolation to convert the discrete coordinates of the joints and the relative Euler angles of the finger bones at the 192 time-instances during one flapping cycle from the videogrammetry measurements into a continuous variation. The space-time continuous variations are then used to generate the kinematics of the skeletal structure at the much finer time-steps (4000 per cycle) used in the simulations. The membrane is affixed to the skeleton and the skeleton provides the displacement boundary conditions for the membrane at these points of union between the membrane and the skeleton.

We note that in the video sequence that forms the basis for the current simulations, the nominal forward velocity of the bat was $3.71$ m/s, with a deceleration of about -1.05 m/s$^2$. This deceleration has implications for the aerodynamic forces on the wing\added{, such as the fact that drag on the wing should exceed the thrust generated by the wing}, and these will be addressed later in the paper. The root chord of the wing ($C$) is 17 cm and the fully stretched wing span for one wing is 36 cm. The Strouhal number\added{$(=fH/U_\infty)$} based on the forward velocity\added{($U_\infty$)} and the peak-to-peak wing tip stroke amplitude\added{($H$)} of 18 cm, is 0.25. The Reynolds number based on the root wing chord and the forward velocity ($Re=\rho U_{\infty}C/\mu$) is 63,070.

\subsection{Flow Simulation}
In this work, direct numerical simulations (DNS) of the flow are performed by solving the incompressible Navier-Stokes equation:
\begin{equation}
    \frac{\partial u_i}{\partial x_i} = 0
\end{equation}
\begin{equation}
    \frac{\partial u_i}{\partial t} + \frac{\partial u_j u_i}{\partial x_j} = -\frac{\partial p}{\partial x_i} + \frac{1}{\text{Re}} \frac{\partial^2 u_i}{\partial x_j x_j}
\end{equation}
where $u_i$ is the velocity and $p$ is the pressure. The sharp-interface immersed boundary solver ViCar3D \citep{mittal2008versatile} is used to simulate this flow. The method allows for the simulation of fluid flow around complex moving bodies on nonconformal Cartesian grids. The Navier-Stokes equations are discretized in space and time using the $2^{nd}$-order finite-difference schemes. The equations are integrated in time using the fractional step method, where the Navier-Stokes equation is split into an advection-diffusion equation and a pressure Poisson equation. In these simulations, the advection-diffusion equation is solved using a line successive-over-relaxation solver, and the pressure Poisson equation is solved using a biconjugate-gradient stabilized solver with scheduled-relaxed Jacobi solver \citep{yang2014acceleration} as a preconditioner. The code has been validated for a variety of computational fluid dynamics studies of flying \citep{zheng2013multi,zhang2015centripetal} and swimming organisms \citep{dong2010computational,seo2022improved}, along with various internal \citep{seo2014effect,bailoor2021computational,zhu2022computational} and external flows \citep{shoele2014computational,shoele2016flutter} interacting with flexible structures.

We note at the outset that while the actual Reynolds number for a bat in forward flight is about 63,000 and it is infeasible to resolve such a high Reynolds number flow with a complex moving/deforming surface, as is the case here. A large body of research on flapping wings and fins \citep{anderson1998oscillating,zheng2013comparative,sekhar2019canonical} has shown that among the various non-dimensional parameters for such flows, the Strouhal number dominates the flow physics. Thus, as long as the Strouhal number is matched and the Reynolds number is sufficiently high so that the boundary layers formed on the flapping control surface are thin compared to the overall movement of the control surface, there is an expectation that the resulting flow physics will be a reasonable approximation to the flow at the full-scale Reynolds number. In the current study, we therefore maintain the Reynolds number at a value of 1000 and ensure that the flow is well resolved by the grid employed.

\begin{figure}
    \centering
    \subfloat[]{\includegraphics[width=0.49\textwidth]{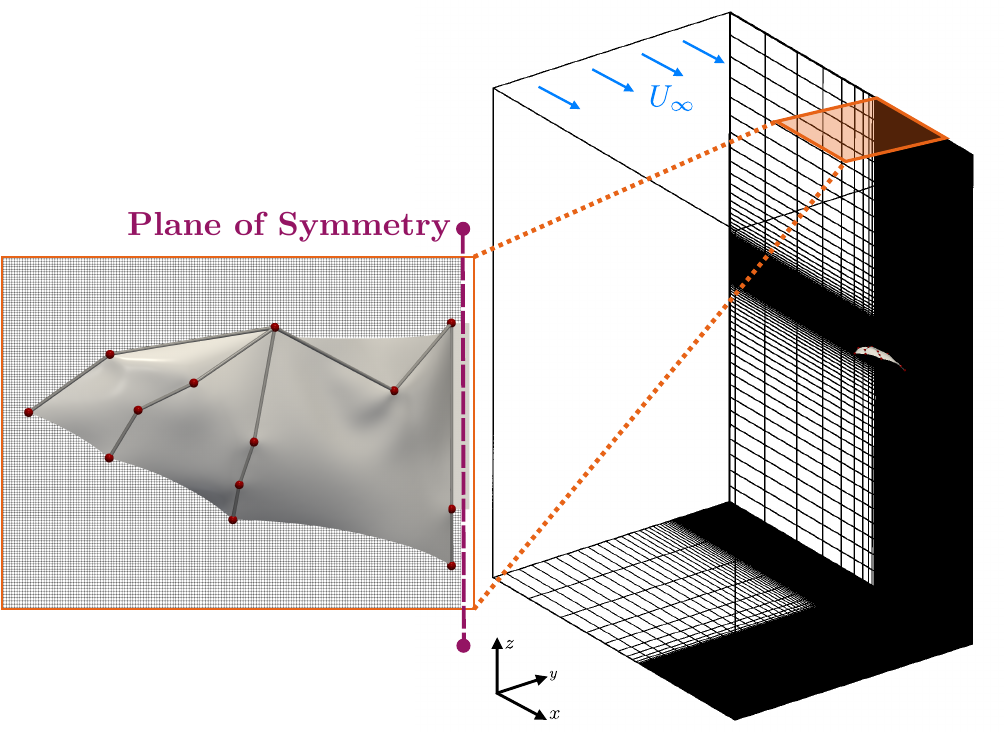}}
    \subfloat[]{\includegraphics[width=0.49\textwidth]{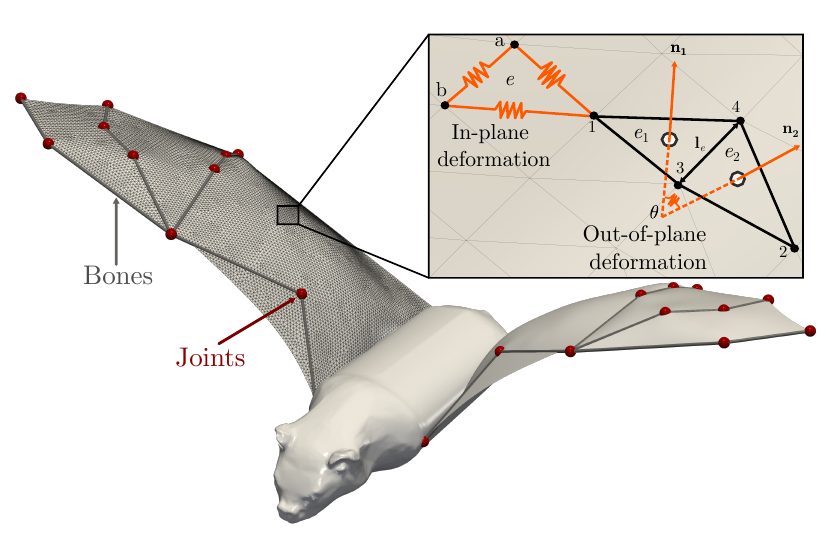}}
    \caption{Visual Representation of Numerical Model. (a) Schematic of the computational domain for flow simulation showing the cartesian grid with immersed bat wing membrane. (b) Representation of bat wing using spring network for structural simulations.}
    \label{fig:computational_domain}
\end{figure}

The computational domain and Cartesian grid used for simulation is shown in figure \ref{fig:computational_domain}(a). A uniform grid is employed in a cuboidal region around the body to resolve the boundary layers and vortex structures, and the grid is stretched away outside this region to the outer boundary. As seen in figure \ref{fig:computational_domain}(a), we are simulating only one wing, and this arrangement could result in the formation of an unphysical root vortex, which could affect the flow over the wing. This issue is mitigated by applying symmetry boundary conditions at the wing root. Thus, the flow mimics the flow over a symmetric dual wing configuration without the body. The nominal grid size for these simulations is $432\times408\times296$ (about 52 million points), which corresponds to a resolution of about 200 points along the wing root chord. We employ a time-step that corresponds to about 4000 time-steps per flapping cycle. We have conducted grid refinement studies to ensure that the time-varying forces generated by the simulated wing are grid-converged (See appendix). Each simulation is carried out for three flapping cycles and the results for the 3rd cycle are used for all the analysis. 
%

\subsubsection{Membrane Dynamics Model and FSI Coupling}
The traditional methods used in simulating structural dynamics are based on finite-element methods \citep{hsu2014fluid,li2019novel}. These methods provide the highest fidelity and have become a gold standard for simulating flexible structures. However, these methods often require a complex computational infrastructure when coupled with immersed boundary methods for FSI simulation \citep{nitti2020immersed}. Moreover, these complex methods require many parameters for the material properties and might not be essential to simulate the dominant dynamics of bat wings, where the structures are thin and slender \citep{spandan2017parallel}. The spring network model provides a good alternative to structural dynamics models for a thin membrane.
This model was initially developed to simulate cloth deformations in computer animations and movies \citep{terzopoulos1988modeling,guendelman2005coupling,bridson2005simulation}. Later, this method was reformulated to perform physics-based simulations like parachutes \citep{kim2013simulation}, flexible revolving wings \citep{truong2020mass}, and very extensively by Verzicco and co-workers \citep{spandan2017parallel, viola2020fluid, verzicco2021collision, viola2022fsei}. The key advantage of this model is its simplicity. The model needs only a small number of parameters to describe the materials, and the governing equation is a very straight-forward Newtonian dynamics equation for point masses. 
Thus, this method is employed in the current study for its robustness and simplicity in coupling with an IBM-based fluid flow solver \citep{marco2016moving}.

In this approach, the wing is modeled as a zero-thickness shell, and the surface is divided into triangular elements that are capable of supporting in-plane as well as bending deformation. Springs attached to the edges of each triangular element control the in-plane/elastic deformation, and bending springs attached to each pair of adjacent triangles control the out-of-plane/bending deformation. As shown in figure \ref{fig:computational_domain}(b), this forms a network of springs, controlling the deformation due to the wings' interaction with the incoming fluid flow.

The dynamics of the nodes of the triangular elements that make up the membrane surface mesh are governed by Newton's $2^\text{nd}$ law, defined for each node in equation \ref{eq:ode}.
\begin{equation}
    \label{eq:ode}
    m \frac{d^2 \mathbf{x}(t)}{dt^2} = \mathbf{f}(\mathbf{x}),
\end{equation}

Here, $m$ is the mass of the node, $\mathbf{x}(t) \in \mathbb{R}^3$ is the instantaneous position of the node, and $\mathbf{f}(\mathbf{x})\in \mathbb{R}^3$ is the force exerted on the node. The total force is a combination of external aerodynamic forces ($\mathbf{f}_\text{ext}$), internal forces ($\mathbf{f}_\text{int}$) induced by springs, and the viscoelastic damping ($\mathbf{f}_\text{damp}$). Both ($\mathbf{f}_\text{int}$) and ($\mathbf{f}_\text{damp}$) would induce a restrictive action on the node's displacement. We can then assemble the system for the surface mesh with $N$ nodes and $E$ triangular elements as
\begin{equation}
    \label{bigode}
    M \frac{d^2 \mathbf{X}(t)}{dt^2} = \mathbf{F}_\text{ext}-\mathbf{F}_\text{int} - \zeta\frac{d \ \mathbf{X}(t)}{d t},
\end{equation}
Here, $M\in \mathbb{R}^{N\times N}$ is the diagonal matrix containing the mass of all the nodes, $\mathbf{X}(t)\in \mathbb{R}^{N\times3}$ are the instantaneous node positions. The external and internal forces induced on the system are then defined as $\mathbf{F}_\text{ext},\mathbf{F}_\text{int} \in \mathbb{R}^{N\times3}$. The last term in equation \ref{bigode} is the expression for the damping term, with $\zeta$ being the structural damping coefficient.

We now focus on calculating the $\mathbf{F}_\text{ext}$ and $\mathbf{F}_\text{int}$. Since we use the same surface mesh for the flow and membrane solver, we can utilize the available fluid stresses at the centroid of the triangular element to compute $\mathbf{F}_\text{ext}$. This is described by equation \ref{fext} where the force on the node $i$ can be collected from $n_e$ elements to which that node is connected. $A^e$ is the triangular area of each element. Here, we consider both the pressure ($\Delta p \mathbf{n} \in \mathbb{R}^{E\times3}$) and the shear ($\Delta \tau \cdot \mathbf{n} \in \mathbb{R}^{E\times3})$ contributions where the $\Delta$ operator refer to the difference across the zero-thickness membrane.
\begin{align}
    \label{fext}
    \mathbf{F}_{\text{ext},i} = \sum_{j=1}^{n_e} \frac{1}{3} \left( -\Delta p \mathbf{I} \mathbf + \Delta\mathbf{\tau}\right)\cdot \mathbf{n}  A^e_j
\end{align}
The internal forces ($\mathbf{F}_\text{int}$) can be computed as a combination of forces induced by elastic springs ($\mathbf{F}_{\text{elas}}$) and bending springs ($\mathbf{F}_{\text{bend}}$). The expression of elastic force induced by a spring attached to an edge between $a$ and $b$ of element $e$ (see figure \ref{fig:computational_domain}(b)) is given by equation \ref{eq:elasticForce} \citep{marco2016moving} where $\mathbf{l}_{e} = \mathbf{x}_{a} - \mathbf{x}_b$.
\begin{align}
    \mathbf{F}^a_\text{elas} = -k_e(|\mathbf{l}_e|-|\mathbf{l}_e|_{t=0})\frac{\mathbf{l}_{e}}{|\mathbf{l}_e|}; 
    \mathbf{F}^b_\text{elas} = -k_e(|\mathbf{l}_e|-|\mathbf{l}_e|_{t=0})\frac{-\mathbf{l}_{e}}{|\mathbf{l}_e|}
    \label{eq:elasticForce}
\end{align}
Next, the internal forces exerted by the bending springs on the nodes of the triangular mesh can be computed from the Helfrich energy approach \citep{fedosov2010systematic,janvcigova2020spring}. The expression for $\mathbf{F}_{\text{bend}}$ is given by equation \ref{eq:bendingForce} where we utilize subscripts 1,2,3 and 4 to refer to the four nodes of two adjacent elements $e_1$ \& $e_2$ with surface normals $\mathbf{n_1}\added{(=(\mathbf{x_1} - \mathbf{x_3})\times(\mathbf{x_1} - \mathbf{x_4}))}$ and $\mathbf{n_2}\added{(=(\mathbf{x_2} - \mathbf{x_3})\times(\mathbf{x_2} - \mathbf{x_4}))}$ respectively which can be viewed in figure \ref{fig:computational_domain}(b).
\begin{align}
    \mathbf{F}^i_\text{bend} = k_b \frac{|\mathbf{l}_e|^2}{|\mathbf{n}_1|+|\mathbf{n}_2|}\Bigg[\sin\frac{\theta}{2} - \sin \frac{\theta_0}{2} \Bigg]s_i
    \label{eq:bendingForce}
\end{align}
where 
\begin{gather}
    s_1 = |\mathbf{l}_e|\frac{\mathbf{n}_1}{|\mathbf{n}_1|^2}; 
    s_2 = |\mathbf{l}_e|\frac{\mathbf{n}_2}{|\mathbf{n}_2|^2}\\
    s_3 = \frac{\left(\mathbf{x}_1-\mathbf{x}_4\right) \cdot \mathbf{l}_e}{|\mathbf{l}_e|} \frac{\mathbf{n}_1}{\left|\mathbf{n}_1\right|^2}+\frac{\left(\mathbf{x}_2-\mathbf{x}_4\right) \cdot \mathbf{l}_e}{|\mathbf{l}_e|} \frac{\mathbf{n}_2}{|\mathbf{n}_2|^2} \\
    s_4 = -\frac{\left(\mathbf{x}_1-\mathbf{x}_3\right) \cdot \mathbf{l}_e}{|\mathbf{l}_e|} \frac{\mathbf{n}_1}{|\mathbf{n}_1|^2}-\frac{\left(\mathbf{x}_2-\mathbf{x}_3\right) \cdot \mathbf{l}_e}{|\mathbf{n}_2|} \frac{\mathbf{n}_2}{|\mathbf{n}_2|^2}
\end{gather}  
The spring constants used in the formulation of elastic ($k_e$) and bending ($k_b$) forces can be computed using
\begin{align}
    k_e = \frac{E^*h^* \left( \Sigma_i A_i^e \right) }{|\mathbf{l}_e|^2}; k_b = \frac{\sqrt{3}E^*h^{*^3}}{12(1-\nu^2)}
\end{align}
In the above equations, $\Sigma_i A_i^e$ is the summation of areas of triangular elements with the common edge for which the spring constant is computed, and both $k_e$ and $k_b$ are calculated for all the edges of the surface mesh. The non-dimensional material properties of the membrane are approximated from \cite{swartz1996mechanical} and are as follows: membrane (solid) to fluid density ratio $\rho^*(={\rho_s}/{\rho_f}) = 1000$; elastic modulus $E^* (={E}/{\rho U_\infty^2}) = 10^6$; membrane thickness is chosen as $h^*(=h/A) = 0.01$ to represent the very thin membrane of the bat wing, and the Poisson's ratio $\nu=0.4$. The damping coefficient $C$ is not known for bat wings, and we chose an intermediate value of $\zeta(=C/2\sqrt{k_b m})=1$. This value should allow for membrane oscillations while at the same time, diminishing numerical instabilities in the membrane.  Equation \ref{bigode} is discretized using the Newmark scheme and an explicit (sequential) coupling between the flow and structural is employed wherein the flow solvers passes the surface pressure and shear \added{force from the previous timestep} to the structural solver, and the structural solver passes the velocity of the surface back to the flow solver. Explicit coupling is fast, robust, and accurate for large solid-to-fluid density ratios \citep{Zheng2010VF}, and verification of this FSI solver is presented in Appendix \ref{app:benchmark}.


\subsubsection{``Bat-inspired'' versus Bat Wing}
Despite the many realistic features of the bat wing that we include in our model, we refer to our wing as ``bat-inspired'' because there are several features that we do not match, and which could potentially have an impact on the dynamics and performance of the wing. As pointed out earlier, the skeletal structure of the hand-wing is made up of slender bones which undergo deformation as well, but this is ignored in our model. Second, the wing membrane is not a passive elastic structure but is comprised of muscle fibers \citep{cheney2022bats,lauber2023rapid}. It is generally understood that these muscle fibers can be activated to change the tension and stiffness of the wing during the flapping cycle \citep{cheney2022bats}. We are unable to model this muscle activation firstly due to the complexity it would entail and secondly, due to the fact that modeling this would require information on the muscle physiology and activation that is not available to us. While this might diminish the direct applicability of the results to bat flight, this simplification is relevant to bat-inspired flapping wing vehicles \citep{ramezani2017biomimetic,zhang2022bio} which typically employ passive (nonactivated) membranes. The thickness of the membrane is also known to vary across the wing \citep{makanya2007structural,lauber2023rapid}, but this feature is not included in our model in order to retain simplicity. \added{Biological membranes are known to have viscoelastic damping, but data for parameterizing this damping is not available to us.} We also do not include a realistic attachment of the wing to the bat's body since this information is not available in the experimental dataset we have employed for the simulations. Based on previous work on bird flight \citep{wang2019computational}, this might be an interesting aspect to explore in the future. Finally, as noted earlier, the Reynolds number in our simulation is about an order-of-magnitude lower than for the actual bat in flight.

\section{Results}\label{sec:results}
We start by describing the results regarding the deformation and dynamics of the membrane and this is followed by a detailed description of the flow features and flow physics associated with the generation of aerodynamic forces on the wing.

\subsection{Membrane Structural Dynamics}
 \begin{figure}
    \centering
    \includegraphics[width=0.75\textwidth]{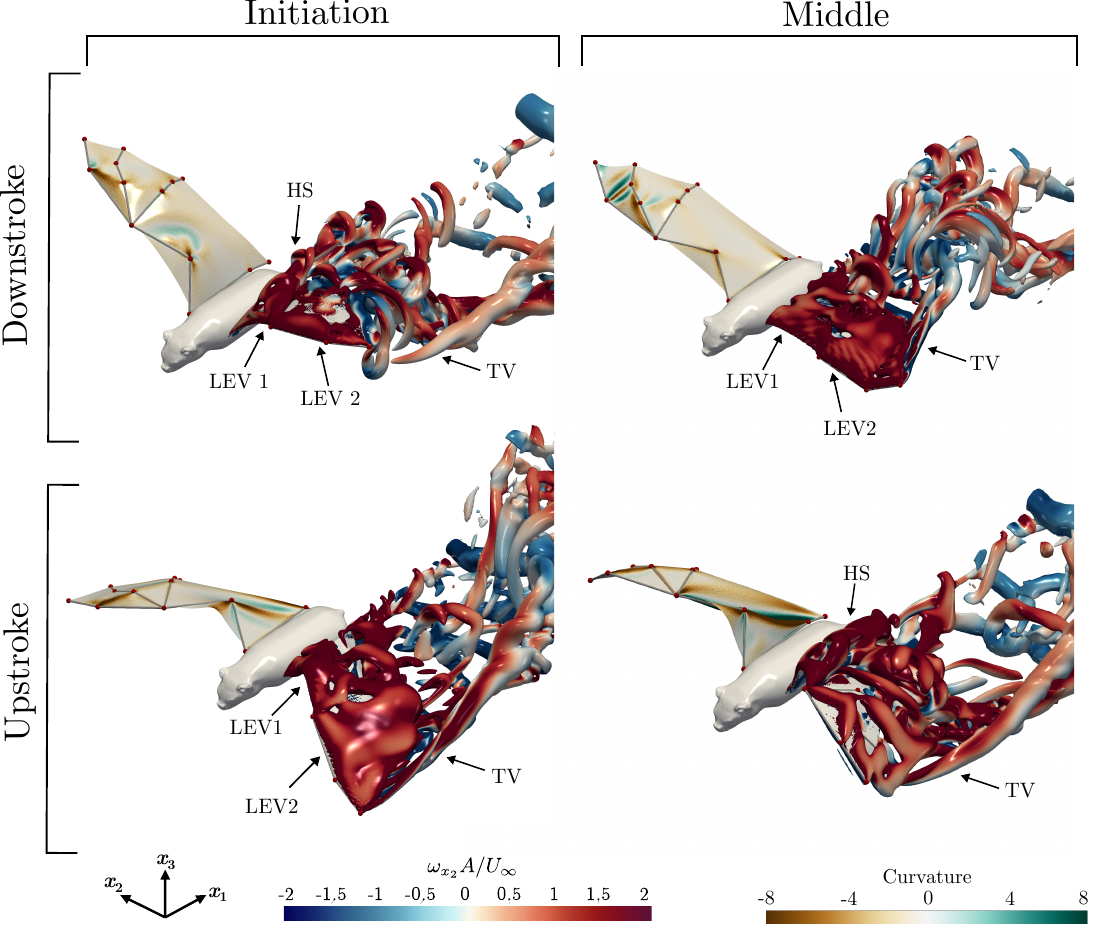}
    \caption{Vortex structure over the (left) wing during the flapping cycle shows via isosurfaces of $Q$\added{-criterion(as defined in eq. \ref{Qcriteqn})} colored by spanwise vorticity. The simulations were performed using the left wing only, and the body and the right wing were added only to facilitate visualization and discussion. The right wing in these plots is used to simultaneously show contours of local wing curvature.}
    \label{fig:QIsoVort}
\end{figure}
\begin{figure}
    \centering
    \subfloat[]{\includegraphics[width=0.48\textwidth]{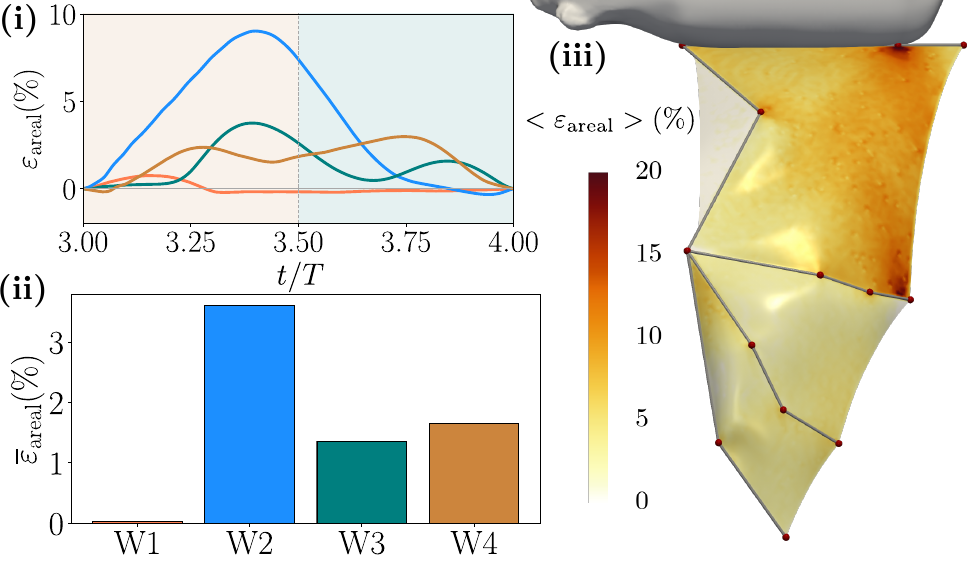}}
  \hfill
    \subfloat[]{\includegraphics[width=0.45\textwidth]{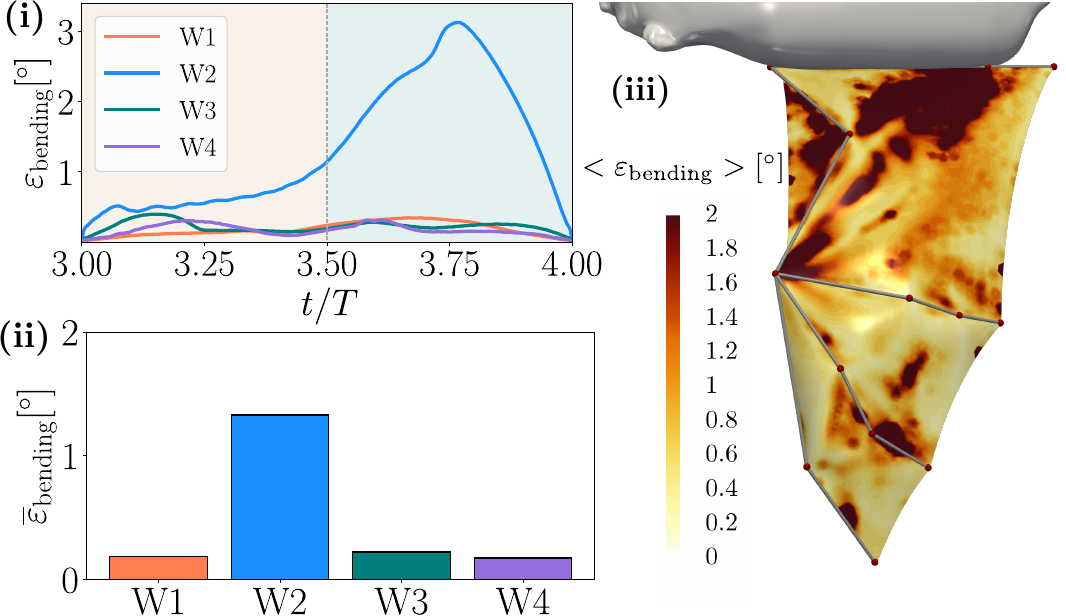}}
    \caption{Time averaged elastic deformation in the wing. (a) Areal strain. (b) Magnitude of bending strain}
    \label{fig:wingSecStrain}
\end{figure}

Figure \ref{fig:QIsoVort} shows a series of snapshots initiating with the bat performing the downstroke, from $t/T$ = 0 to 0.5, followed by the upstroke or recovery stroke, from $t/T$ = 0.5 to 1. In this and some other plots, we show the body and the two wings in order to facilitate the discussion. The contours on the right wing (facing forward) show values of local membrane curvature and the structures on the left wing are vortex structures (to be discussed in a later section). These figures clearly show the complex geometric deformation experienced by the wing as it is articulated by the movement of the digits and as the membrane undergoes additional flow-induced deformation. Due to highly complex kinematics, different wing regions experience different levels of deformation. Figure \ref{fig:wingSecStrain} shows the time-varying and cycle-averaged normalized areal strain ($\varepsilon_\text{areal}$) and the bending strain ($\varepsilon_\text{bending}$) for the wing membrane. The areal strain for the i$^\text{th}$ triangular element is defined as   
\begin{align}
    \varepsilon^{(i)}_{\text{areal}} (t) = \frac{A^{(i)}(t) - A^{(i)}(t=0)}{A^{(i)}(t=0)}
    \label{areasStrain}
\end{align}
where $A^{(i)}$ is the element area. The bending strain is defined for the $j^\text{th}$ \added{triangular element} as
\begin{align}
    \added{\varepsilon^{(j)}_{\text{bending}} (t) = \text{cos}^{-1}( \mathbf{n}^{(j)}(t) \cdot \mathbf{n}^{(j)}(t=0))}
    \label{bendingStrain}
\end{align}
where \added{$\mathbf{n}^{j}(t)$ is the normal} to the \deleted{at the }$j^\text{th}$  \deleted{between the two neighboring} triangular element\deleted{s} as shown in figure \ref{fig:computational_domain}(b).

We note that the plagiopatagium (W2) experiences the highest areal strain with a time-averaged value of about 3.7\% as shown in figure \ref{fig:wingSecStrain}(a)(ii). The areal strain increases as the wing performs the downstroke, with the maximum strain (up to about 8.8\%) generated towards the end of the downstroke. We also show the distribution of the time-averaged areal strain over the surface in figure \ref{fig:wingSecStrain}(a)(iii) and note that the regions towards the trailing-edge experience the highest strain. The areal strains for the dactylopatagium medius and major are roughly of the same magnitude and more evenly distributed over the entire flapping cycle. 

Figure. \ref{fig:wingSecStrain}(b)(i) shows the time variation of the bending strains for the different segments of the wing, and figure \ref{fig:wingSecStrain}(b)(iii) shows the contours of the time-averaged local bending strain for the wing and the segment average values are presented in the figure. \ref{fig:wingSecStrain}(b)(ii). We note that while the propatagium (W1) experiences very little areal strain, it undergoes significant geometric deformation during the flapping cycle. The plagiopatagium exhibits significant bending strain, but unlike the areal strain that peaks during the downstroke, the bending strain for this segment peaks during the mid-upstroke, when the retraction of the digits generates slack in the membrane and allows inertia and aerodynamic forces to generate geometric deformation.

\begin{figure}
    \centering
    \subfloat[ ]{\includegraphics[width=0.46\textwidth]{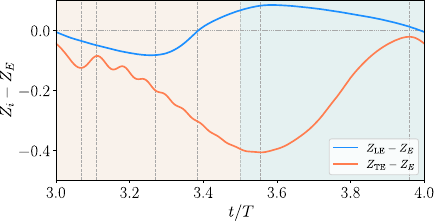}}
    \hfill
    \subfloat[ ]{\includegraphics[width=0.53\textwidth]{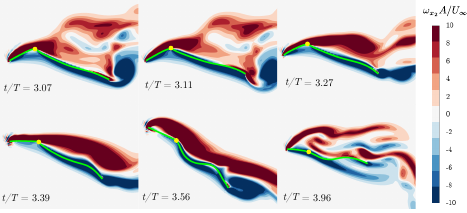}}
    \caption{Time variation of the movement of the proximal sections of the wing. (a) Relative vertical distance between the elbow joint ($Z_E$) and the propatagium leading-edge ($Z_{LE}$) and the plagiopatagium trailing rdge ($Z_{TE}$). (b) Contours of spanwise vorticity at a section passing through the middle of the propatagium and the plagiopatagium. \added{Here the deformed wing is marked with a green curve and elbow joint is marked with yellow circle.}}
    \label{fig:flutter}
\end{figure}

Figure \ref{fig:flutter}(a) shows the time variation of the relative vertical distance between the elbow joint and a point at the LE and the TE along the center of the propatagium and figure \ref{fig:flutter}(b) shows snapshots of the wing chord at a spanwise location that cuts through the center of the propatagium.  Together, these two plots show the appearance of several interesting aeroelastic phenomena. First the time variation of $(Z_\text{TE} - Z_E)$ shows oscillations, which indicate the presence of a flutter instability in the plagiopatagium (W2), during the downstroke. The implication of these on the aerodynamic forces will be discussed later in the paper. On the upstroke, the chordwise sectional plots also show undulations in the plagiopatagium (W2) which appear due to the slack in the membrane as the digits are retracted during the upstroke. 

The propatagium undergoes significant bending deformation as well. During the downstroke, it has a downward inclination and this enables the shear layer to stay attached to the propatagium during the downstroke. During the upstroke, the propatagium ``buckles'' upwards due to the retraction of the digits, and this reduces the downward angle of the leading edge and helps maintain the attachment of the shear layer on the dorsal surface of the propatagium. As we will show later, despite the relatively small area of this segment, its location at the leading-edge coupled with its complex deformation, gives this segment an out-sized role in the generation of aerodynamic forces. 

\subsection{Vortical Features of the Flow}
In this section, we will describe the vortex structures that are generated as the wing interacts with the flow. As before, the flapping cycle initiates with the bat performing the downstroke, from $t/T$ = 0 to 0.5, followed by the upstroke or the recovery stroke, from $t/T = $0.5 to 1, and figure \ref{fig:QIsoVort} shows snapshots of the vortex structures generated at a few key phases in the cycle. The downstroke starts with the formation of leading-edge vortices along the leading edges of propatagium (LEV 1) and dactylopatagium medius (LEV 2) due to the interaction of incoming airflow and the downward-moving leading-edge of the membrane wing. The LEV over the propatagium sheds periodically during the downstroke forming a series of horseshoe (HS) shaped vortices. The tip vortex (TV) that formed during the previous cycle is seen detaching from the wingtip. The vortices originating from propatagium detach from the wing and merge with the trailing edge vortices, giving rise to complex wake vortices. While the LEV at the propatagium continuously forms and breaks, the LEV at the dactylopatagium major and medius remains attached. This is due to reduced local effective angle-of-attack due to the supination of this part of the wing. This LEV transitions to a triangular shaped vortex and spans the dactylopatagium medius and major. The triangular vortex remains attached over the entire downstroke and first quarter of the upstroke. The downward wing motion also results in the formation of a strong tip vortex (TV) that is identifiable throughout the flapping cycle.

\subsection{Aerodynamic Forces}
\begin{figure}
    \centering
    \subfloat[]{\includegraphics[width=0.475\textwidth]{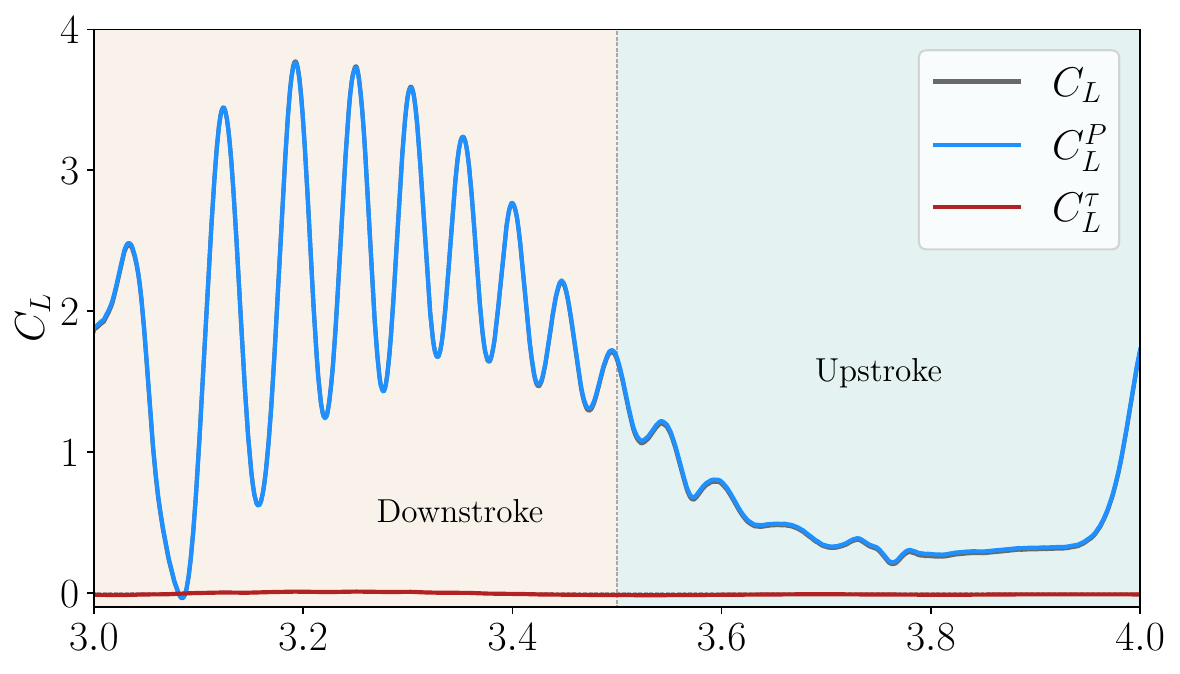}}
    \hfill
    \subfloat[]{\includegraphics[width=0.50\textwidth]{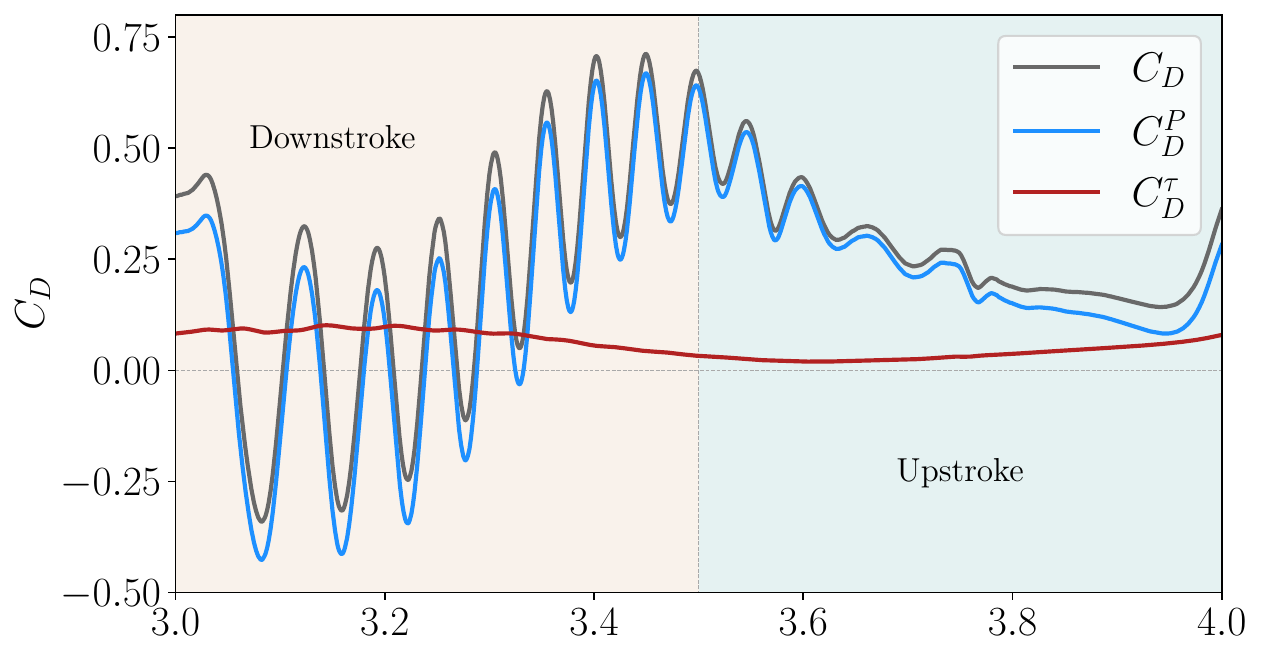}}
    \caption{Time variation of the aerodynamic forces over one flapping cycle and decomposition into contributions from pressure and shear effects. (a) Lift (b) drag.}
    \label{fig:aeroforces}
\end{figure}
\begin{figure}
    \centering
    \subfloat[]{\includegraphics[width=0.45\textwidth]{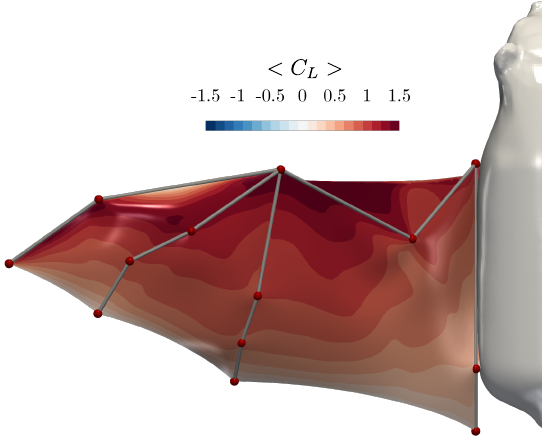}}
    \hfill
    \subfloat[]{\includegraphics[width=0.45\textwidth]{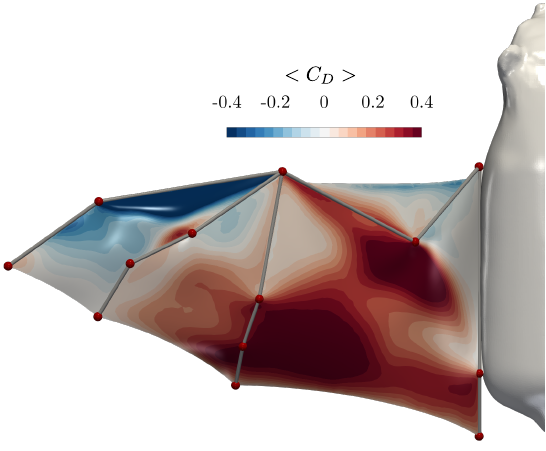}}
    \caption{Surface distribution of the time-averaged area density of the force coefficients. (a) Lift, (b) drag.  }
    \label{fig:aeroforcesSurface}
\end{figure}

We focus here on the generation of lift and drag forces and quantify them in terms of the lift and drag coefficients which are defined as follows
\begin{align}
    \label{forceCoeffDef}
    C_L = \frac{L}{1/2 \rho U_\infty^2 A_w} ; C_D = \frac{D}{1/2 \rho U_\infty^2 A_w}
\end{align}
where $L$ and $D$ are the lift and drag forces respectively, and $A_w$ is the area of the wing \added{in the initial undeformed state}. The time traces of the aerodynamic force coefficient over one flapping cycle are shown in figure \ref{fig:aeroforces} and Table \ref{tab:coefficients} presents the cycle-averaged force coefficients. First, we note that as expected, the shear stresses make a negligible contributions to the lift and a small (24\%) contribution to the mean drag at this relatively high Reynolds number. We will therefore focus on the pressure component of the aerodynamic loads in the rest of the paper. We note that the majority of the lift is generated during the downstroke. However, the wing also generates a non-negligible \emph{positive} lift during the upstroke. Indeed, the lift generated during the upstroke is $\approx25\%$ of the downstroke lift. In a later section, we will explore the origin of this positive lift during the upstroke using force partitioning. The drag reaches large positive values at the two ends of the stoke but the wing actually generates a negative drag (i.e. thrust) during the middle of the downstroke as the wing pitches downwards and the pressure difference across the propatagium (see figure \ref{fig:aeroforcesSurface}) generated a component of force in the upstream direction. Both the lift and drag force exhibit an oscillatory behavior which is particularly prominent in the downstroke, and this is connected with the flutter in the plagiopatagium noted earlier. We note that flutter oscillations were also observed in the coupled FSI simulations by \cite{joshi2020variational}. In the next section, we will examine the fluid-dynamic origin of these force oscillations using force partitioning.
\begin{table}
  \begin{center}
\def~{\hphantom{0}}
  \begin{tabular}{lccc|ccc}
      \hline
      \hline
      \textbf{Force Type} & \multicolumn{3}{c|}{\textbf{$\overline{C}_L$}} & \multicolumn{3}{c}{\textbf{$\overline{C}_D$}} \\
      \cline{2-7}
       & \textbf{Downstroke} & \textbf{Upstroke} & \textbf{Total} & \textbf{Downstroke} & \textbf{Upstroke} & \textbf{Total} \\
      \hline
      \textbf{Total} & 2.04 & 0.55 & 1.30 & 0.22 & 0.27 & 0.25 \\
      \textbf{Pressure Force} & 2.04 & 0.54 & 1.29 & 0.14 & 0.23 & 0.19 \\
      \quad\textit{Vortex-Induced Force} & 1.97 & 0.42 & 1.20 & 0.21 & 0.21 & 0.21 \\
      \quad\textit{Kinematic Force} & -0.09 & 0.06 & -0.01 & -0.04 & -0.01 & -0.02 \\
      \quad\textit{Viscous Diffusion Force} & 0.17 & 0.06 & 0.10 & -0.02 & 0.03 & 0.005 \\
      \textbf{Shear Force} & 0.00 & 0.01 & 0.01 & 0.08 & 0.04 & 0.06 \\
      \hline
      \hline
  \end{tabular}
  \caption{Cycle-averaged values of the coefficients of lift ($\overline{C}_L$) and drag ($\overline{C}_D$) experienced by the wing and their various components.}
  \label{tab:coefficients}
  \end{center}
\end{table}

In figure \ref{fig:aeroforcesSurface}, we plot the area-variation of the area density (i.e. force coefficient per unit area) of the pressure component of the time-average of the local lift and drag coefficients indicated with the $<\cdot>$ operator on the surface of the wing. We note here that this force coefficient density is indicative of the effectiveness of a localized region to generate the particular force in question and  these plots reveal the following interesting characteristics:
\begin{enumerate}
\item they indicate that the effectiveness of lift, drag and thrust generation varies quite significantly over the wing.
\item The regions close to the leading edge are most effective at generating lift. Indeed, the propatagium is particularly effective in generating lift and this due to the formation of the LEV over this region of the wing as shown before. The effectiveness of lift generation decreases as we move away from the leading-edge towards the trailing edge. 
\item The plagiopatagium segment of the wing is most effective in generating drag. This drag is mostly generated at the end of the downstroke where the wing has a large angle-of-attack and the flow is separated over this portion of the wing (see figure corresponding to $t/T$ = 3.39 in figure \ref{fig:flutter}). The trailing-edge region of the dactylopatagium major (W3) close to the plagiopatagium is also an effective drag generator for the same reason.
\item The dactylopatagium medius (W4), especially its leading-edge region, is singularly effective in generating thrust, and this is also due to the formation of the LEV during the downstroke when the ventral surface of this segment pitches down and the suction pressure on this segment generated a pressure force in the thrust direction.
\item The leading-edge region of the propatagium also generates a small amount of thrust, and this is also connected with its pitch down orientation during the downstroke. Thus, the propatagium is a small but important component of the flight apparatus.
\end{enumerate}
\begin{figure}
    \centering
    \subfloat[]{\includegraphics[width=0.49\textwidth]{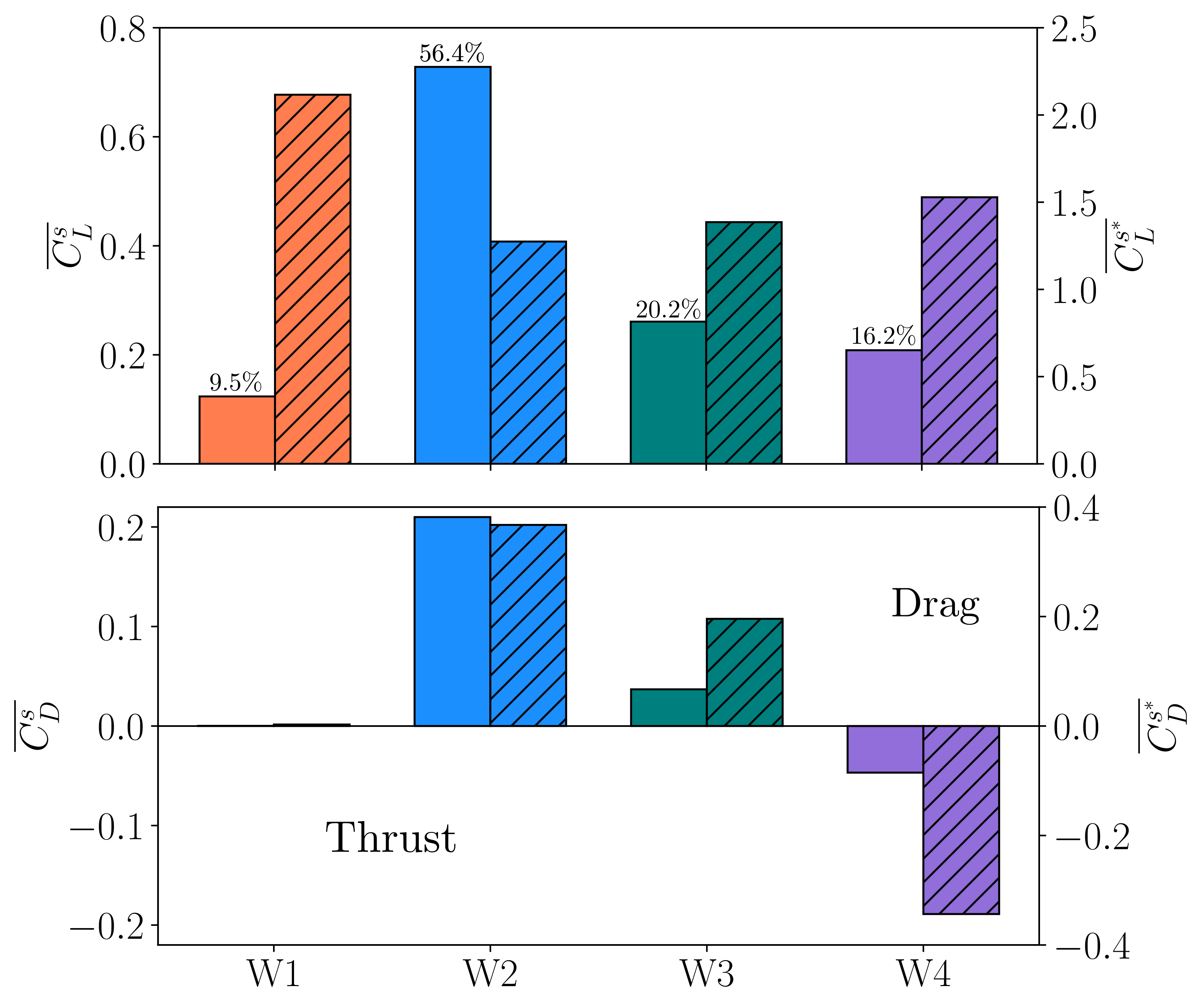}}
    \hfill
    \subfloat[]{\includegraphics[width=0.46\textwidth]{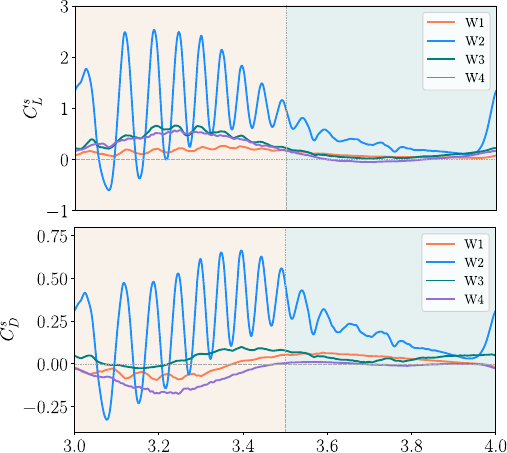}}
    \caption{Contribution to pressure lift and drag forces from different segments of the wing. (a) presents time-averaged segmental force coefficient with left $y$-label indicating values for nominal coefficients corresponding to filled bars and right $y$-label indicating values for area normalized coefficient corresponding to dashed bars. (b) presents time-varying nominal force coefficients.}
    \label{fig:wingSecLift}
\end{figure}

In order to further distinguish the contributions of the different wing segments on the aerodynamic lift we quantify the net force contributions from each of the four wing segments. The first set of plots (in figure \ref{fig:wingSecLift}) is for the time and space averaged net lift force generated by each segment presented as a coefficient, i.e. $C^{(s)}_L = L^{(s)}/({0.5\rho U_\infty^2A_w})$, where $s=1,...,4$ corresponds to the four segments of the wing membrane. These plots shows that the plagiopatagium (W2) is the primary generator of the lift force on the wing (it generated 56\% of the total mean lift). This is firstly because of the large area of this segment (it constitutes 57\% of the total area of the wing). Second, this region of the wing undergoes relatively small movement during flapping due to its attachment to the body. Thus, this portion of the wing acts more like a static foil during forward flight and generates lift due to its positive angle-of-attack through the entire flapping stroke (see figure \ref{fig:flutter}). Next in terms of their contributions are the dactylopatagium major (W3) and the dactylopatagium medius (W4), and the smallest contribution (about 9\% of the total lift), comes from the propatagium. These relative contributions are directly consistent with the total areas of each of these segments. The total lift generated by the armwing (W1+W2) is about 31\% more than the handwing (W3+W4) and is consistent with the observation made by \cite{fan2022power} who employed a reduced-order-model to obtain vertical force estimates for bat of different species flying at similar flight speed. We note that \cite{lauber2023rapid} excluded the proximal regions of the bat wing in their simulations, and this could therefore miss the primary contributor to the lift and drag force for these wings. The exclusion of the proximal region of the wing combined with the exclusion of the drag generating plagiopatagium, would also tend to accentuate the thrust generating characteristics of the wing.

We now consider the lift generated by each segment \emph{per unit area} of the segment, i.e. $C^{{(s)*}}_L = L^{(s)}/({0.5\rho U_\infty^2A^{(s)}})$. This quantity, shown as hatched bars in figure \ref{fig:wingSecLift}(a), is indicative of how effective a given segment is in generating lift, and here we find that the propatagium has a value of $\overline{C^{s*}_L}$ equal to 2.1, whereas the other three segments have values ranging from 1.27 to 1.53. Thus, per unit area, the propatagium is far more effective (approximately 50\% more effective) in generating lift than the other segments of the wing. As pointed out earlier, this is due to the unique deformation profile of this segment of the wing membrane during the flapping cycle\added{, as well as its placement near the leading-edge of the wing.}

Figure \ref{fig:wingSecLift}(b) shows the time variation of this segmental lift coefficient, and we note the large oscillations in the lift force for the plagiopatagium during the downstroke. These are connected with the flutter oscillations that occur for this segment of the wing. The lift from the plagiopatagium reaches a maximum during mid-downstroke. To a large extent, this is due to the fact that this corresponds to the largest downward velocity of this segment and also a point in time when the areal strain is close to its maximum. The force coefficient corresponding to the segments of the handwing also shows the oscillation connected with some flutter in these segments. These results are consistent with the results from the modeling of the handwing by  \cite{lauber2023rapid}.

The lower plots in figure \ref{fig:wingSecLift} show the corresponding data for the drag force. The key observations are that firstly, the vast majority of the drag is generated by the plagiopatagium, and this drag peaks during the end of the downstroke. Second, the dactylopatagium medius generates thrust per unit area far out of proportion to its net thrust indicating the role that unsteady effects associated with the large scale flapping and pitching play in generating the aerodynamic loads on this segment of the wing. These results are also consistent with the horizontal forces obtained for a bat flying at similar flight speed by \cite{fan2022power} where the armwing is found to dominate over the handwing in terms of drag. \cite{fan2022power} observed a very minimal drag contribution from the handwing, whereas we find that this region of the wing generates a small but measurable thrust.

\subsection{Insights from Force Partitioning}
The wing stroke of bats gives rise to complex vortex structures as shown above, and these have significant effects on the induced pressure forces on the wing. The acceleration, rotation, deformation and flutter of the wing membrane during the flapping stroke could also induce forces on the wing through added-mass effects. In order to gain an understanding of the contributions of these different features and mechanisms, we employ the force partitioning method \citep{zhang2015centripetal,menon2022contribution,seo2022improved}. The force partitioning method (FPM) employs an influence field ($\phi_i({\bf x})$) which is calculated separately for drag ($\phi_1$) and lift ($\phi_3$) at any given time-instance by solving a Laplace equation with the boundary shape at that time-instance (see appendix \ref{app:fpm}). By projecting the Navier-Stokes equations onto the gradient of the influence field, the pressure forces are partitioned into contributions from what we term as the ``kinematic force'' that depends on the acceleration \added{(both linear acceleration as well as the centripetal acceleration - see \cite{zhang2015centripetal})} of the immersed surface (${F_i^\kappa=-\rho \int_B \hat{n}.\left( {dU_B}/{dt} \right)\phi_idS}$), the ``vortex-induced force (VIF)'' (${F_i^Q = -2 \rho \int_{V_f} Q\phi_i dV}$) and the component due to the viscous diffusion of momentum ($F_i^\mu=\mu \int_{V_f} (\nabla^2 u).\nabla \phi_i dV$). In these expressions, $U_B$ is the velocity of the segment $dS$ on the body $B$, and $V_f$ is the fluid volume.  Finally, $Q$ is the observable used to identify vortices in a flow and is related to the velocity field as follows:
\begin{equation}
Q=\frac{1}{2} \left( \lVert {\bf \Omega} \rVert^2- \lVert {\bf S} \rVert^2 \right) \equiv -\frac{1}{2}{\bf \nabla} \cdot ({\bf u}\cdot {\bf \nabla u})\, ,
\label{Qcriteqn}
\end{equation}
where $\Omega$ and ${\bf S}$ are the rotation and strain tensors. Some additional details of FPM are included in the appendix \ref{app:fpm} and further details can be found in \cite{zhang2015centripetal,menon2022contribution,seo2022improved}.

\subsubsection{Kinematic versus Vortex-Induced Forces}
\begin{figure}
    \centering
    \subfloat[]{\includegraphics[width=0.47\textwidth]{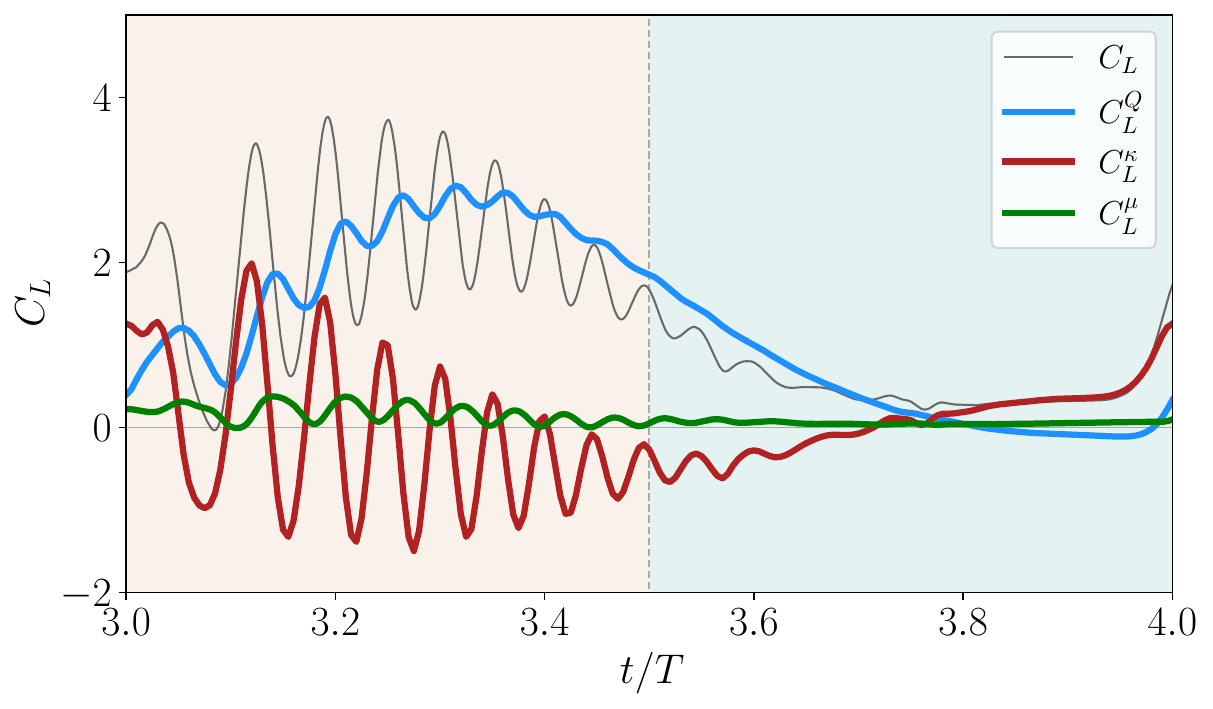}}
    \hfill
    \subfloat[]{\includegraphics[width=0.49\textwidth]{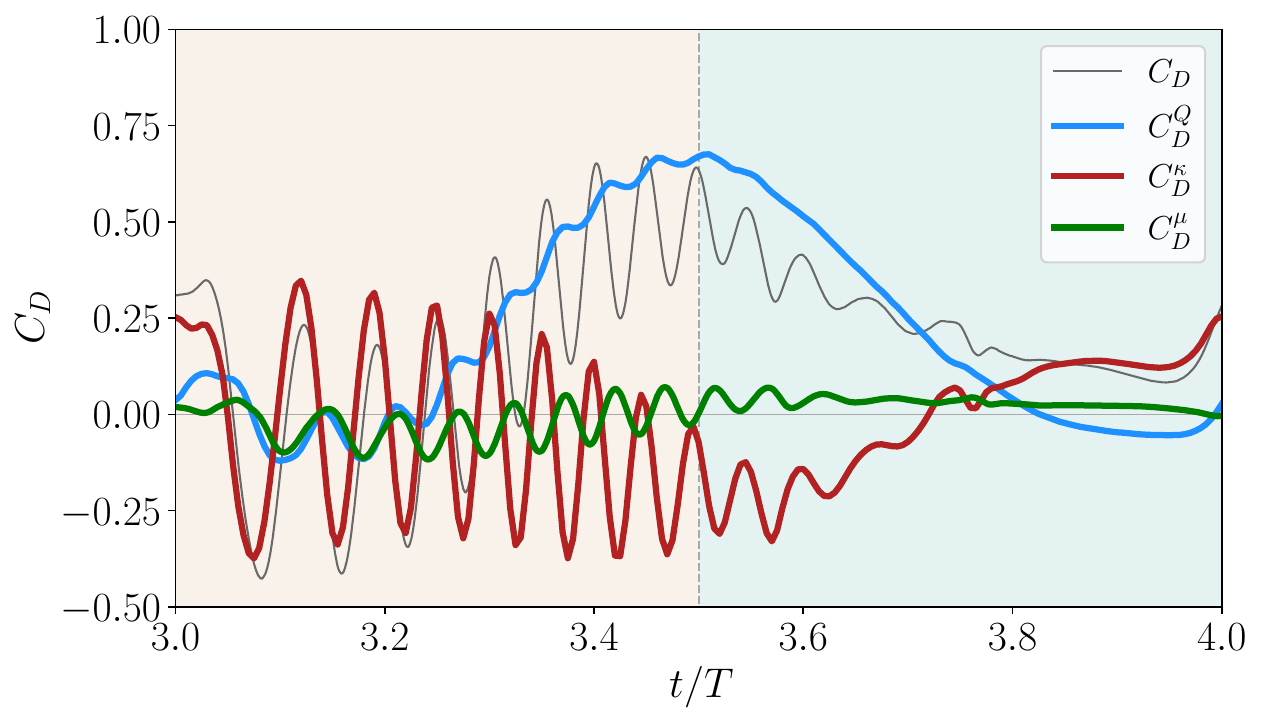}}
    \caption{Decomposition of pressure forces on the wing into kinematic, vortex-induced and viscous diffusion induced partitions using the force partitioning method. (a) pressure lift coefficient $C_L^P$ (b) pressure drag coefficient $C_D^P$}
    \label{fig:forceCurvesFPM}
\end{figure}
Figure \ref{fig:forceCurvesFPM} and table \ref{tab:coefficients} shows the decomposition of the pressure component of lift and drag into its three partitions (kinematic, vortex induced and viscous diffusion). First, we note that the VIF component is responsible for generating the vast majority of net pressure lift and drag forces on the wing, and that the viscous momentum diffusion induced pressure force can be mostly ignored. The kinematic force, which, as shown by \cite{zhang2015centripetal} consists in general of the linear and centripetal acceleration reaction forces, averages to a very small value. Furthermore, the time variation of the various partitions of the pressure-induced lift and drag forces shows that the primary source of oscillation in the forces is the kinematic force. This observation is consistent in the time evolution behaviors of both lift ($C_L^{\kappa}$) and drag ($C_D^{\kappa}$). The oscillations in the kinematic force are associated with the linear acceleration reaction mechanism (a.k.a. added mass effect) associated with the flutter in the plagiopatagium segment of the membrane. The oscillations in the vortex-induced forces ($C_L^{Q}$ \& $C_D^{Q}$) are small compared to the added-mass forces as evident in figure \ref{fig:forceCurvesFPM} and are likely caused by the temporal oscillation in the boundary layers over the wing caused by the flutter. 

Finally, the force partitioning also provides insight into the generation of positive lift during the upstroke. FPM shows that as the wing begins its upstroke and accelerates upwards, the vortex-induced lift begins to reduce in magnitude and is effectively zero in the second half of the upstroke. On the other hand, the kinematic lift is slightly negative in the first part of the upstroke as the wing is accelerating upwards but becomes slightly positive in the second half of the upstroke as the wing undergoes a vertical \emph{deceleration}. Thus, with VIF providing positive lift during the first half and the kinematic force providing positive lift in the second half of the upstroke, the entire upstroke ends up generating positive lift. While the net lift of the upstroke is small (about 25\% of the lift for the downstroke), this could contribute to reducing the vertical oscillations in the body of the bat during flight and facilitate a more steady body posture that is useful for visual navigation and tracking.







\subsubsection{Contributions of Dorsal and Ventral Vortex Structures to the VIF}
Starting with the lift, we note that since $\phi_3$ is determined by the vertical component of the vector normal to (and pointing into) the surface of the wing, $\phi_3$ is negative above the wing (on the dorsal surface) and positive below the wing (on the ventral surface). Thus, the lift contributions of the flow structures on the dorsal and ventral regions of the wing can be further partitioned by separating the VIF integral into fluid volumes with negative $\phi_3$ and positive $\phi_3$, respectively.

\begin{figure}
    \centering
    \begin{subfigure}{0.455\textwidth}
        \centering
        \includegraphics[width=\textwidth]{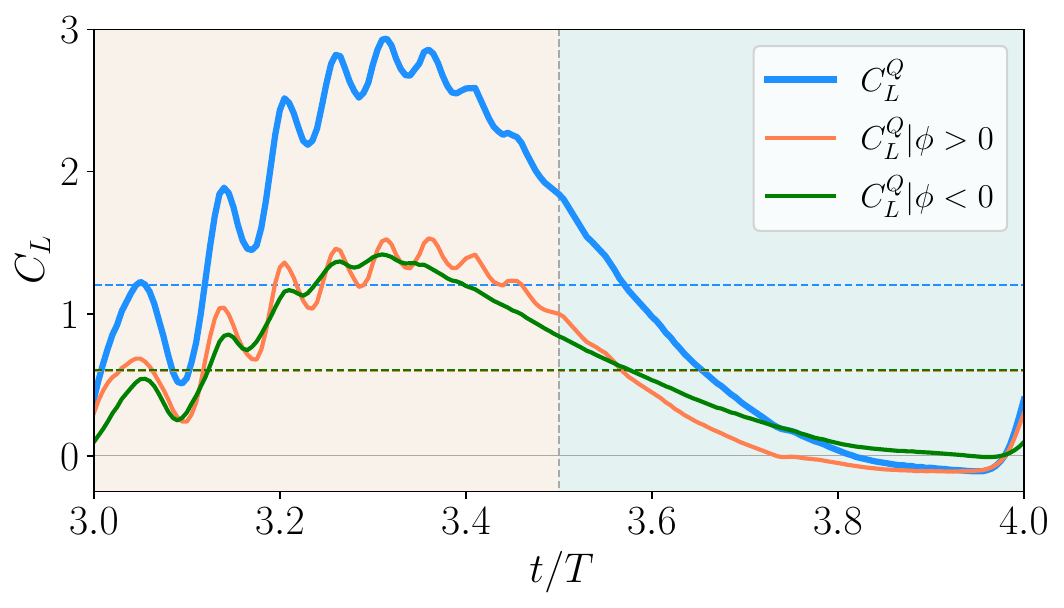}
        \caption{}
    \end{subfigure}
    \hfill
        \begin{subfigure}{0.48\textwidth}
        \centering
        \includegraphics[width=\textwidth]{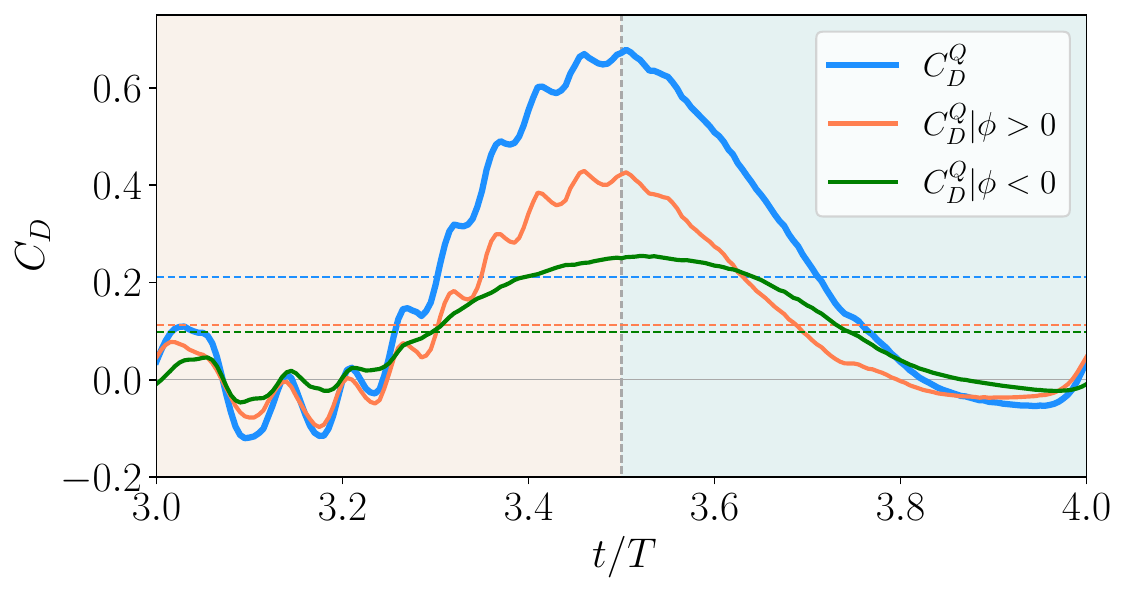}
        \caption{}
    \end{subfigure}
    \caption{Decomposition of vortex-induced forces into contributions from fluid volumes corresponding to +ve $\phi_3$ (ventral) and -ve $\phi_3$ (dorsal)  regions. (a) Lift (b) Drag}
    \label{fig:FPMSurfaceDecomp}
\end{figure}

Figure \ref{fig:FPMSurfaceDecomp}(a) shows the vortex-induced contribution to lift for both surfaces of the wing and we note that both surfaces contribute almost equally to the net lift. The vortex-induced force density is given by $-2Q\phi_3$ and therefore, given the sign of $\phi_3$ on the two surfaces, positive lift generation is expected to be associated with a positive $Q$ (i.e. rotation dominant vortex cores) on the dorsal surface and a negative $Q$ (i.e. strain dominant regions) on the ventral surface of the wing. Thus, the fact that both sides of the wing generate equal magnitudes of lift suggests that the vortices on the suction (dorsal) side of the wing and the shear layers on the pressure (ventral) side of the wing are equally important. This is different from static lifting wings where a significant portion of the lift is associated with the leading-edge suction \citep{abbott2012theory}. This is also distinct from flapping foils \citep{raut2024hydrodynamic} and fish fins \citep{seo2022improved} where the leading-edge vortex that develops on the suction side of the control surface plays a dominant role in the generation of pressure forces. Figure \ref{fig:FPMSurfaceDecomp}(b) shows the plot corresponding to the dorsal and ventral decomposition of the vortex-induced drag force and while there are some differences in the temporal variation of the drag force, both sides also contribute equally to the total vortex-induced drag.

Finally, it is also worth noting that the ventral contributions of the vortex-induced lift and drag forces exhibit larger flutter-induced oscillations than the dorsal contributions. This corresponds well to the flow physics since the flutter is associated with the plagiopatagium and the flow as well as the vortices over the dorsal surface are separated, and therefore distant from the surface of the plagiopatagium (see figure \ref{fig:flutter}). On the other hand, the shear layers on the ventral surface are attached to the plagiopatagium and, therefore, are more acutely affected by the flutter in the plagiopatagium.

\subsubsection{Correlation of Vortex-Induced Forces to Flow Features}
\begin{figure}
    \centering
    \includegraphics[width=0.7\textwidth]{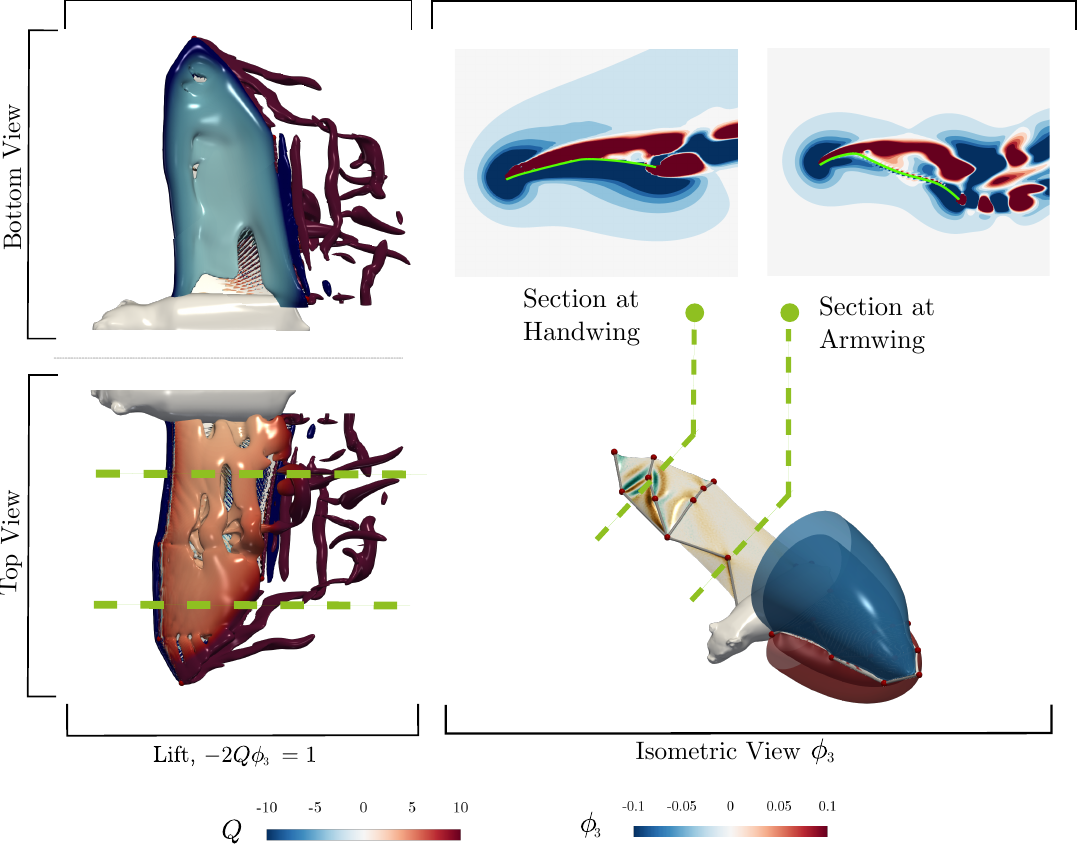}
    \caption{Analysis of vortex-induced lift ($-2Q\phi_3$) using FPM. Here, we plot the isosurfaces of vortex-induced lift colored by $Q$ at $t/T = 3.32$, corresponding to the instance with maximum $-2Q\phi_3$. We also present isosurfaces of $\phi_3$ field along with two slices at armwing and handwing showing contours of $Q$.}
    \label{fig:VIL}
\end{figure}

We now examine the vortex-induced forces in more detail and correlate them to the flow features through the aid of FPM. The peak in the vortex-induced lift occurs close to mid-downstroke and figure \ref{fig:VIL} shows isosurfaces of $\phi_3$ at this phase of the flapping cycle. Figure \ref{fig:VIL} shows isosurfaces corresponding to a positive vortex-induced lift force density $-2Q\phi_3=1$, which is an intermediate value of lift density, colored by the value of $Q$. The top-right of the figure shows the contours of $Q$ at two spanwise cross-sections of the wing at this time instance: one through the arm-wing (center of propatagium and plagiopatagium) and one through the hand-wing, which goes through the dactylopatagium medius and major. As discussed earlier, given the sign of $\phi_3$ on the dorsal and ventral surfaces, we expect that positive lift will be associated with positive(negative) $Q$ on the dorsal(ventral) surfaces, and that is indeed what these figures shows. On the dorsal surface we see that the flow is dominated with a separated shear layer that rolls up into a series of vortices that are aligned with the wing-span and all of these structures contribute positive lift. On the ventral surface, the flow in the hand-wing region is dominated by an attached boundary layer that comprises exclusively of negative values of $Q$ and, therefore, also contributes positive lift to the wing.

We now turn to a similar analysis of the drag force but separately examine the phase in the flapping cycle when the wing generates thrust ($t/T=3.10$) and when it generates maximum drag (end downstroke at $t/T=3.50$). We note that the particular flight configuration chosen here corresponds to the situation when the bat is decelerating. Thus, unlike flight at a steady speed where the wings would generate a net positive thrust to counteract the drag on the body, the wings in our simulation are not expected to generate significant thrust. Nevertheless, our wing does generate thrust at certain stages in the cycle and it is of interest to understand how this is accomplished since this can provide information about the steady flight of bats. 
\begin{figure}
    \centering
    \begin{subfigure}{0.7\textwidth}
        \centering
        \includegraphics[width=\textwidth]{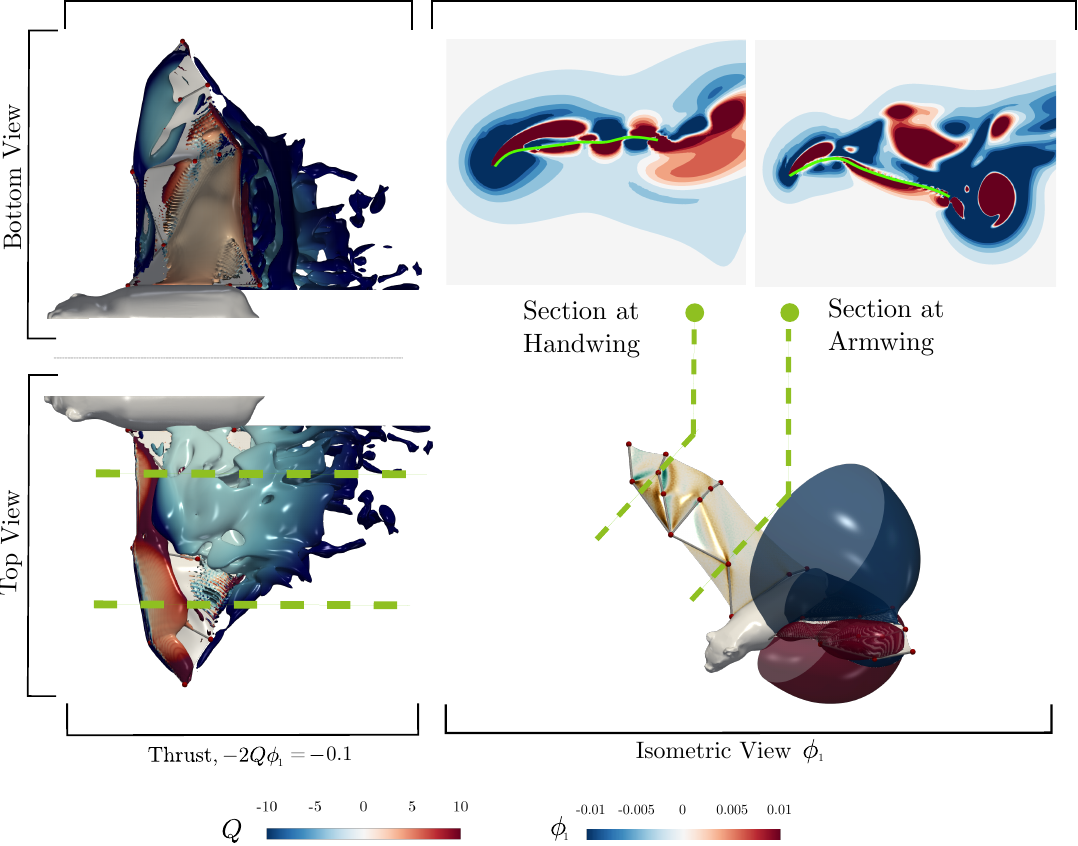}
       \caption{ }
    \end{subfigure} \\
        \begin{subfigure}{0.7\textwidth}
        \centering
        \includegraphics[width=\textwidth]{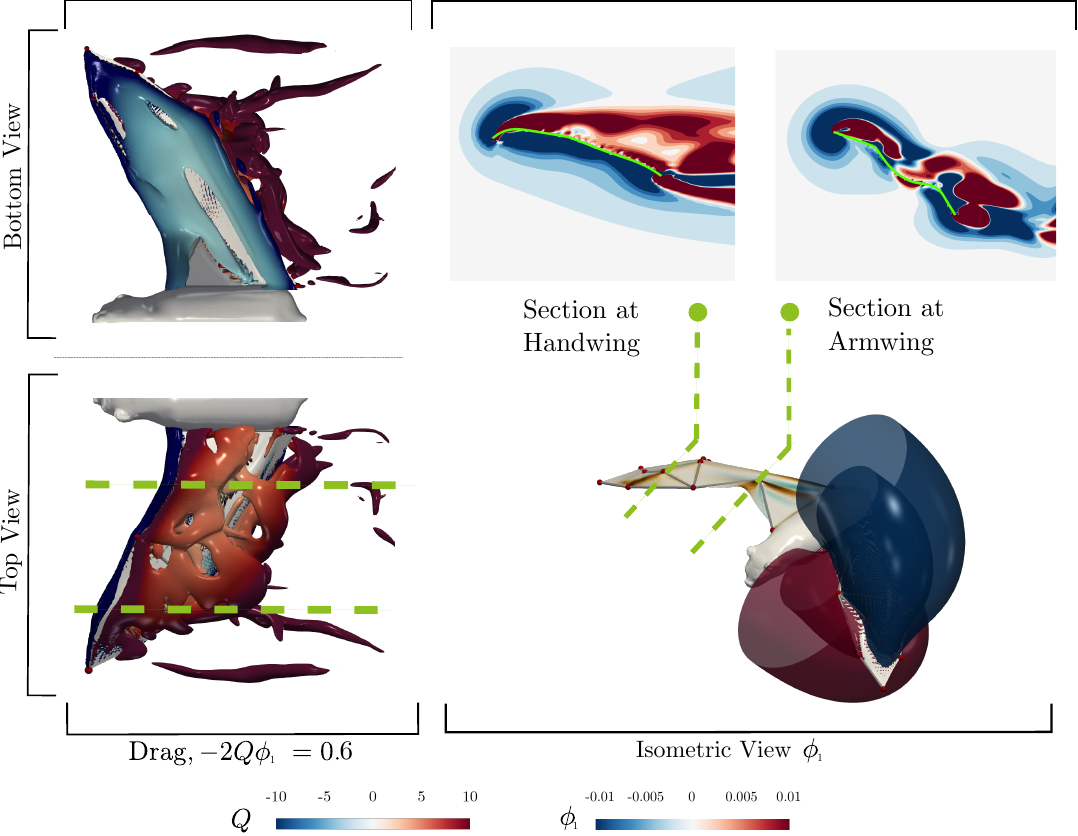}
       \caption{ }
    \end{subfigure}
    \caption{Analysis of vortex-induced horizontal forces ($-2Q\phi_1$) using FPM. Here, we plot the isosurfaces of $-2Q\phi_1$ colored by $Q$ along with isosurfaces of $\phi_1$ and two slices at arm and hand wing showing contours of $Q$. Inset (a) presents results for thrust force at $t/T=3.1$, and inset (b) presents results for drag force at $t/T=3.5$.}
    \label{fig:VID}
\end{figure}

Figure \ref{fig:VID}(b) shows the plots for the maximum thrust (early downstroke at $t/T=3.10$) situation and the plots on the left show isosurfaces of positive thrust force density corresponding to $-2Q\phi_1=-0.1$ colored by the values of $Q$. The view of the dorsal side of the wing shows that thrust generation during early downstroke is the result of several distinct features. The red isosurfaces indicate the presence of an LEV that extends over the LE of the propatagium as well as the dactylopatagium medius, and which generates positive thrust. As visible from the inset cross-sectional plots, the leading-edge of the wing undergoes rapid pronation at the start of the downstroke and this is responsible for the formation of these LEVs as well as for pointing the surface of the wing associated with these vortices into the thrust direction. Indeed, the positive isosurfaces of $\phi_1$ over this region of the dorsal surface clearly highlights its orientation towards the thrust direction. On the ventral surface of the wing, the downwards motion of the wing tip generates a region of strong shear (with a negative $Q$) which coupled with the pronation of the leading edge also generates positive thrust. Finally, there is also a large strain dominated region of vorticity over the dorsal surface of the plagiopatagium that, coupled with the supinated orientation of this wing segment, generates positive thrust. This is associated with the downward induced flow and associated strain generated between the vortices in the separated shear layer. Thus, the leading-edge regions of the propatagium and the dactylopatagium medius, as well as the supination of the plagiopatagium during downstroke could play an important role in generating thrust for supporting steady or accelerating flight in bats.

Finally, we examine the end of the downstroke ($t/T=3.50$) that corresponds to the maximum drag and we plot the isosurfaces of $-2Q\phi_1=0.6$ in figure \ref{fig:VID}(b) to identify structures that contribute substantially to the drag. At this stage in the flapping cycle the wing has already assumed a mostly supinated position along its entire surface. On the dorsal surface, we note several rolled up vortices that generate a suction force that coupled with the supinated orientation of the surface, generate drag. Similarly, the ventral side boundary layer is strain dominated which generates a pressure force onto the ventral surface, thereby also generating a component in the direction of drag. 

\subsection{Comparison with Equivalent non-Articulated Wings}
In the previous sections, we have described the aerodynamics and aero-structural dynamics of the fully articulated, flexible bat-hand wing. The extensive geometric deformation during flapping, including local bending and stretching are features of the bat wing that distinguish it from the wings of insects and birds. In this section we explore the role of this unique feature, i.e. massive wing deformation associated with wing articulation, in the generation of aerodynamic forces by using models with simplified wing kinematics that eliminate this feature while preserving some of the other key aspects of the wing stroke kinematics.

\begin{figure}
    \centering
    \includegraphics[width=0.8\textwidth]{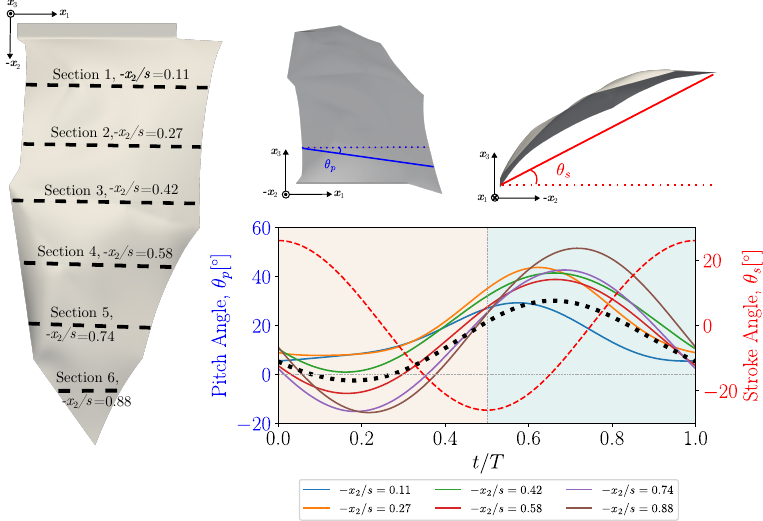}
    \caption{Development of equivalent stiff wings. Here we show the sections along which pitch angles ($\theta_p$) were extracted. The time variation of pitch angles along with stroke angles ($\theta_s$) was used to develop ``twisted wing''. The dashed black line is the span-averaged twist angle, which along with the stroke angles was used to develop ``flat-wing''.}
    \label{fig:StiffConstruct}
\end{figure}

At a fundamental level, flapping wings are control surfaces that undergo periodic and simultaneous stroking (about the wing root) and pitching. Figure \ref{fig:StiffConstruct} shows the pitch angles (measured for a straight chord line joining the leading edge to the trailing-edge) at six locations along the span for the wing obtained from the FSI simulations. Figure \ref{fig:StiffConstruct} shows the corresponding stroke angle for a plane joining the wing root to the wing-tip. These data are used to generate two simplified models of the bat wing kinematics that exclude the large geometric deformation due to the articulation of the finger joints. The first model is a planar ``flat-wing'' undergoing pitching and flapping oscillations. For this flat-wing model, the time varying pitch is the same at every spanwise location and this matches the pitch corresponding to the average pitch at these 6 locations. The time varying stroke angle is matched to that shown in figure \ref{fig:StiffConstruct}. In the second ``twisted-wing'' model, the stroke angle is the same as the flat-wing model but we allow the pitch to vary along the span as well (thereby generating a spanwise twist) and we match (best-fit) this local pitch at the six locations on the span identified above. These wings with simplified kinematics have some similarities to the relatively stiffer flapping wings simulated in previous studies \citep{song2014three,seo2019mechanism,joshi2020full}.

\begin{figure}
    \centering
    \subfloat[]{\includegraphics[width=0.57\textwidth]{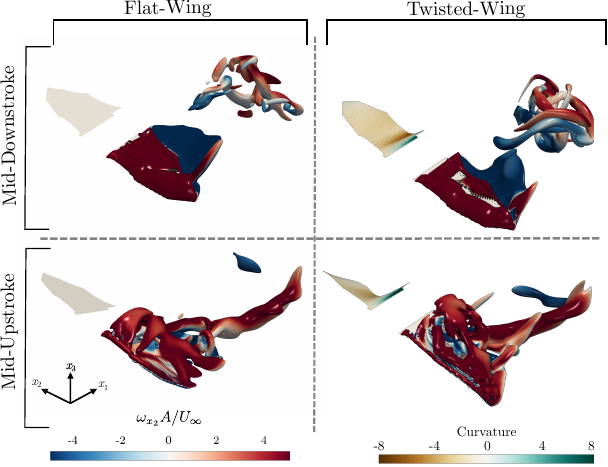}}
    \subfloat[]{\includegraphics[width=0.42\textwidth]{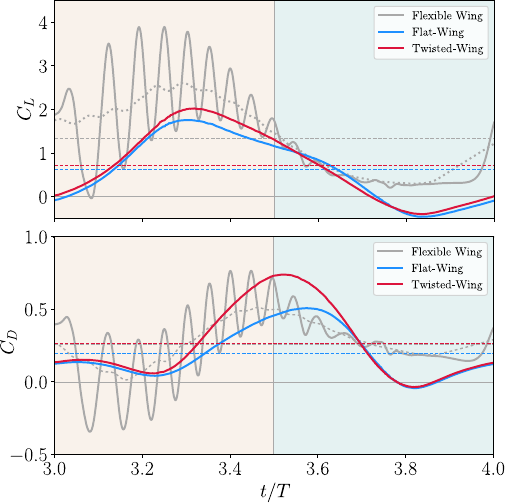}}
    \caption{Simulation results for comparative study. (a) Isosurfaces of $Q$ colored with spanwise vorticity ($\omega_{x_2} A/U_\infty$). (b) Time-variation of lift and drag along with a comparison with the flexible wing.}
    \label{fig:flapIso}
\end{figure}

Figure \ref{fig:flapIso}(a) shows the vortex structures for these two synthesized wings at mid-upstroke and mid-downstroke. We note that these vortex structures are quite different from those observed for the actual bat wing. During the mid-downstroke,  we observe the formation of a leading-edge vortex for both wings. In the case of the twisted-wing, the LEV has a nearly uniform strength across the entire LE, whereas, for the flat-wing, the strength of the vortex increases with the LE edge and is maximum at the tip. For the actual wing, we observed a tip-vortex lasting through the entire downstroke but this is absent in these simulations. Moreover, the roll-up observed during the downstroke due to breakup of LEV in flexible wings is also absent in these cases despite having an equivalent wing planform, flapping angle, and local twist angles, indicating the significant role of the articulation-driven geometric deformation in the actual wing. 

Figure \ref{fig:flapIso}(b) shows the time variation of the lift and drag coefficients for the two wings superposed on the corresponding variation for the actual articulated wing. We note that similar to the articulated wing, major contributions to the lift are generated during the downstroke, although the articulated wing generates a larger lift than these derived wings. Also, unlike the articulated wing, which could sustain positive lift during the entire upstroke, these derived wing with reduced deformation generate negative lift post-mid-upstroke. As we noted earlier, this positive lift during upstroke is connected with the linear acceleration reaction (a.k.a added mass) effect, and this seems to be absent in these derived wings. The differences in the lift profile of the flat and twisted wings are quite small. Overall, the articulated wing generates twice the lift of these derived wings. The time traces of the drag coefficient, indicate that all three wings experience a similar magnitude of mean drag over the flapping cycle and the stiff wings do not generate thrust at any time during the flapping cycle. These comparisons indicate that even with the planform shape and overall stroking-pitching kinematics of the bat wing matched to the bat wing, the exclusion of the geometric deformation that is generated by active wing articulation, greatly diminishes the wing's overall aerodynamic performance.

\section{Conclusions}
A fully coupled fluid-structural computational model is employed to investigate the aero-structural dynamics of a bat wing in a forward flight. We have attempted a high-degree of realism in the model with respect to the wing anatomy and kinematics, as well as the geometric and elastic deformation associated with active articulation and flow-induced deformation. The following are the key findings of the study:
\begin{enumerate}
    \item The areal strain can reach up about 9\%, and is particularly significant in the plagiopatagium, which accounts for the largest area among all the segments of the wing, and which constitutes an important region with respect to the aerodynamic loads. The plagiopatagium also experiences large bending strain, which peaks in the upstroke as the phalanges are retracted during the upstroke resulting in the development of slack in the membranes. The plagiopatagium also experiences aeroelastic flutter during the downstroke.
    \item The propatagium undergoes significant localized bending deformation that results in large supination during the downstroke and large spanwise bending during the upstroke due to the development of slack in the wing membrane. Given that the propatagium is located at the leading-edge, these deformations have important aerodynamic effects.
    \item A regional analysis of the wing shows that the effectiveness of the ability of the wing to generate lift, drag and thrust varies quite significantly over the area of the wing. The leading-edge of the wing, which is covered by the propatagium and the dactylopatagium medius is particularly effective in generating lift. The trailing region of the plagiopatagium is a highly effective generator of drag. Finally, the thrust seems to be primarily generated by the dactylopatagium medius. \added{The latter is because the dactylopatagium medius behaves as a pitching-heaving flapping foil, and such a motion is known to generate thrust.}
    \item The force partitioning method shows that the vast majority of the mean aerodynamic force comes from the vortex-induced component. The kinematic force (associated with the added mass effect) generated a negligible net contribution to the forces but does generate a net positive lift over the upstroke.
    \item The force partitioning also suggests that the vortex structure on the dorsal side and the ventral are equally important for aerodynamic force generation. 
    \item The force partitioning also enables us to understand how the movement of the wing coupled with its local inclination couples with the vortex structures to generate lift, drag, and thrust at different phases in the flapping cycle.
    \item Comparison of the fully articulated against with simpler wings with significantly reduced geometric deformation derived from the available kinematics provide some indication of the advantages of wing articulation.
\end{enumerate}

In summary, the current study, despite its limitations, provides several new insights regarding the dynamics and aerodynamics of flight with bat-inspired membrane wings. In particular, we have highlighted the important role that the deformation of the different segments of the elastic membrane wing plays in the generation of aerodynamic lift, thrust and drag forces. We expect that these insights will provide a better understanding of the biomechanics and locomotory capabilities of these animals and will also be useful in the development of bio-inspired flying vehicles. The study has also demonstrated the capabilities of the new FSI model, which we expect will find use for investigating a variety of flow problems involving complex membraneous structures. Finally, the study also demonstrates the usefulness of the force partitioning method for enabling insights into the causality of pressure force in this highly complex vortex-dominated flow.


\backsection[Acknowledgments]{The authors would like to acknowledge Professor Sharon Schwartz and Dr. Dan Riskin for providing the data for the bat wing kinematics from their experiments. We would also like to acknowledge Professor Marco de Tullio for insightful discussions regarding the spring-mass membrane model.}

\backsection[Funding]{This research was supported by the US National Science Foundation through Grant No. CBET-2011619. The development of the FSI solver also benefited from US Office of Naval Research Grant N00014-22-1-2655. \added{The work benefited from the Rockfish HPC system at Advanced Research Computing at Hopkins(ARCH).}}

\backsection[Declaration of interests]{The authors report no conflict of interest.}




\appendix

\section{\label{app:fpm} The Force Partitioning Method}
We provide a brief summary of the force partitioning method here and the reader is referred to earlier papers  \citep{zhang2015centripetal,menon2021quantitative,menon2022contribution,seo2022improved} for more details. The pressure force is the dominant component of the total aerodynamic force experienced by immersed bodies/surfaces such as the bat wing. Pressure is an elliptic variable simultaneously affected by various physical mechanisms/features such as vortices, kinematics of the immersed body, acceleration of freestream flow, and viscous diffusion. The fluid-structure interaction of bats is the result of strongly coupled and complex physics, and dissecting the total aerodynamic force into these mechanisms is necessary to understand the force generation mechanism. The force partitioning method allows the decomposition of the pressure forces by first computing an influence field, which is obtained as the solution of the following Laplace equation
\begin{align}
    \nabla^2\phi_i = 0, \quad \text{in $V_f$   with   } \hat{n}.\nabla \phi_i =\Bigg\{
    \begin{split}
        & n_i, \text{on $B$} \\ 
        & 0,   \text{ on $\Sigma$}
    \end{split}
    \quad \text{for $i=1,2,3.$}
    \label{phi_equation}
\end{align}
where, $V_f$ is the fluid volume, $n_i$ is component of the surface normal on the body $B$ in the direction of the force being partitioned, and $\Sigma$ is the outer boundary. Thus, the influence field at any given time, depends on the shape of the immersed boundary at that time and in the current study, the influence field is also solved for using the sharp-interface immersed boundary solver. 

Projection of the Navier-Stokes equation onto the gradient of the influence field results in the following partitioning:
\begin{align}
\begin{split}
    F_i =\int_B p n_i dS =  &\overbrace{-\rho \int_B \hat{n}.\left({dU_B}/{dt} \right) \phi_idS}^\text{$F^\kappa$} \overbrace{-2 \rho \int_{V_f} Q\phi_i dV}^\text{$F^Q$} + \overbrace{\mu \int_{V_f} (\nabla^2 u).\nabla \phi_i dV}^\text{$F^\mu$} \\
    &\overbrace{- \rho \int_{\Sigma} \hat{n}.\Bigg(\frac{du}{dt} \phi_i\Bigg) dS}^\text{$F^\Sigma$} \text{ for $i = 1,2,3$}    
\end{split}
\label{fpm_decomposition}
\end{align}
where $F^\kappa$, $F^Q$, $F^\mu$, and $F^\Sigma$ are the kinematics-induced (associated with \added{linear and centripetal} acceleration reaction mechanisms \citep{zhang2015centripetal}), vortex-induced, viscous momentum diffusion-induced and the outer boundary induced partitions of the pressure force. For a case where the outer boundary is such that the material acceleration of the flow is small (such as the current case), the outer boundary contribution to the pressure force can be neglected. 

\section{Grid Independence\label{grid_dom_inde_section}}
\label{app:grid}
\begin{figure}
    \centering
    \subfloat[]{\includegraphics[width=0.49\textwidth]{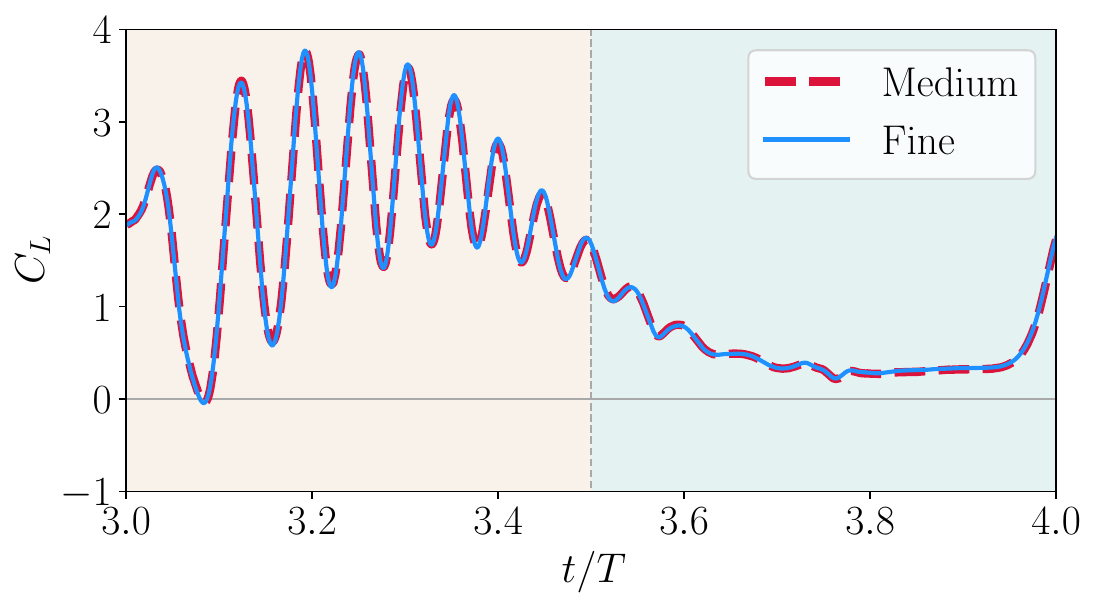}}
    \subfloat[]{\includegraphics[width=0.49\textwidth]{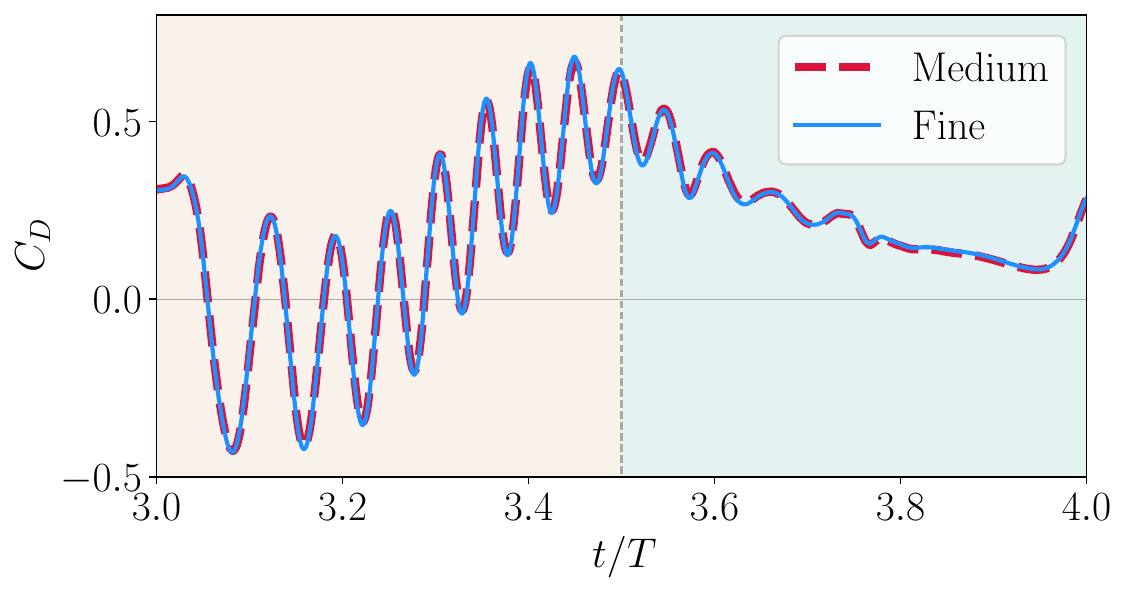}}
    \caption{Results from grid convergence analysis. Time variation of force coefficients for two distinct grids (a) $C_L$ (b) $C_D$}
    \label{grid_inde}
\end{figure}

We demonstrate the grid independence of the results by simulating on two grids - the ``fine'' mesh, which is the mesh used for the study and a ``medium'' mesh where the resolution in the region around the wing is reduced by a factor of two in each of the three directions. The fine grid has a total of 52 million grid points and the coarse mesh has 22 million points. The resolution of the surface discretization in the structural model is also decreased nominally by a factor of two for the coarser mesh. Both simulations are run for three cycles and the time-varying lift and drag coefficients compared (see figure \ref{grid_inde}) to assess grid convergence. 
The mean values of the lift and drag coefficients show a difference of 0.8\% and 0.3\% respectively, and the difference in the corresponding root-mean-square values is 0.57\% and 0.27\%. These data indicate that the fine mesh provide adequate for these simulations.

\section{Benchmarking the Flow-Membrane Interaction Solver}
\begin{figure}
    \centering
    \subfloat[]{\includegraphics[width=0.49\textwidth]{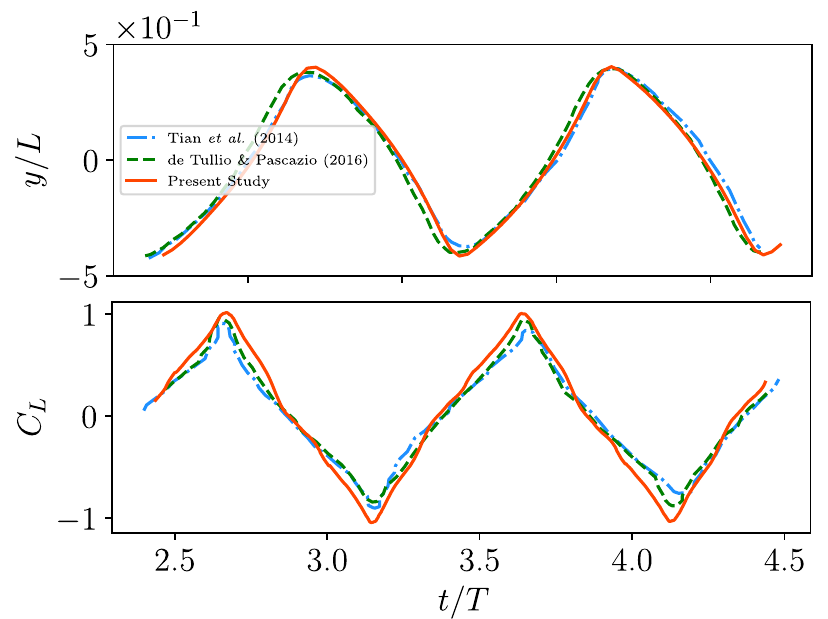}}
    \hfill
    \subfloat[]{\includegraphics[width=0.45\textwidth]{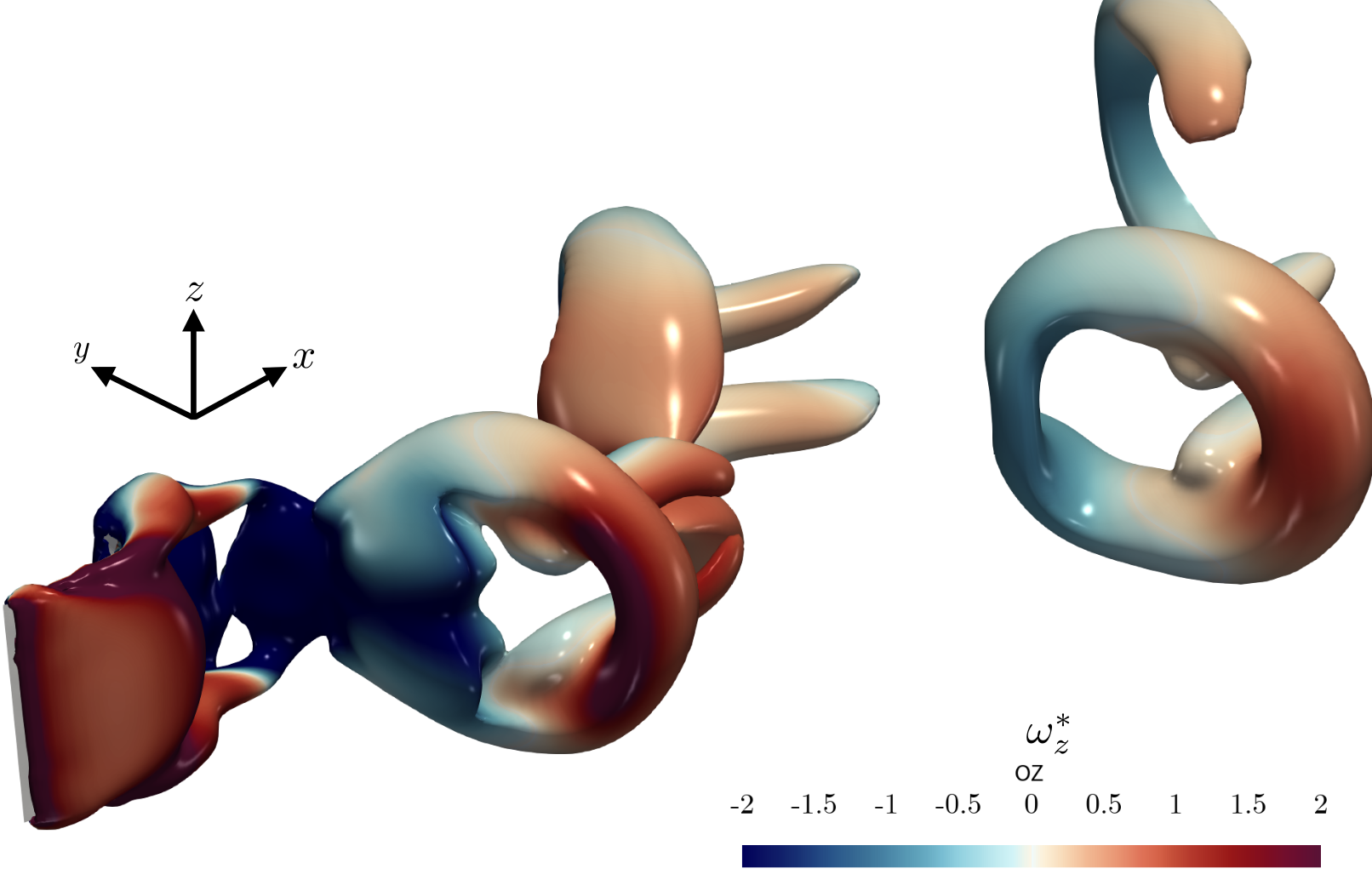}}
    \caption{Quantitative and qualitative results from the benchmark study for the flow-induced flapping of a flag. (a) Comparison of (top) the displacement of mid-point $y/L$ at the trailing edge of the membrane and (bottom) the lateral force coefficient $C_L$ (b) Membrane deformation and flow structures identified using the isosurfaces of $Q$-criterion and colored by spanwise vorticity $\omega_z^*$. }
    \label{benchmark_fsi}
\end{figure}

To benchmark the FSI solver, we employ the case of the 3D flag in a flow that has been the subjects of several previous computational modeling studies with different methods \citep{tian2014fluid,marco2016moving}. The flag in question here is square in shape and undergoes flow-induced flutter as it interacts with the flow. The simulations are performed at a Reynolds number (based on the incoming velocity and flag dimension $L$) of 200 and with a non-dimensional thickness $h/L=0.01$, bending rigidity $k_b = 0.0001$ and mass ratio ${\rho_s h}/{\rho_f L} = 1$. The comparison is made against the works of \cite{tian2014fluid} and \cite{marco2016moving}. \cite{tian2014fluid} utilized a finite-element method based membrane model whereas \cite{marco2016moving} utilized a spring-network model similar to ours but with a different mathematical formulation. Figure \ref{benchmark_fsi} shows the qualitative and quantitative results from our simulation compared to these previous studies. Figure \ref{benchmark_fsi}(a) shows a good match in the time-variation of lateral force and displacement of the midpoint at the trailing edge obtained from our simulation with those reported in the literature. In the accompanying table, we compare the trailing edge displacement amplitude $A/L$ and Strouhal number St with the ones reported in the literature and find that the match is reasonably good. The results proved the ability of our model to perform accurate fluid-structure interaction simulations of membranes subject to extensional and bending strain.
\label{app:benchmark}
\begin{table}
  \begin{center}
\def~{\hphantom{0}}
  \begin{tabular}{lccc}
      & \textbf{Present Study} & \textbf{\cite{tian2014fluid}} & \textbf{\cite{marco2016moving}} \\
      \hline
      $A/L$ & 0.819 & 0.812 & 0.795 \\
      St  & 0.275 & 0.263 & 0.265 \\
  \end{tabular}
  \caption{Comparison of key quantities for benchmarking the FSI solver}
  \label{tab_benchmark}
  \end{center}
\end{table}

\bibliographystyle{jfm}
\bibliography{bat_paper}

\end{document}